\documentclass[sort,numbers,journal=jacsat,manuscript=article]{achemso}
\SectionNumbersOn
\usepackage{tikz} %Format quantum circuits
\usetikzlibrary{quantikz2}%Format quantum circuits

\usepackage{multicol}
\usepackage{graphicx}% Include figure files
\usepackage{dcolumn}% Align table columns on decimal point
\usepackage{braket}
\usepackage{bm}% bold math
\usepackage{pgffor}
\usepackage[utf8]{inputenc}
\usepackage[T1]{fontenc}
\usepackage{mathptmx}
\usepackage{listings}
\usepackage{rotating} % Rotating table
\usepackage{caption}
\usepackage{color}
\usepackage{multirow} % for table cells to span rows
\usepackage{pifont} % for checkmarks
\usepackage{epsfig}
\usepackage{amsmath} % matrix
\usepackage{float}
\usepackage{booktabs}
\usepackage{tabularx}
\usepackage{natbib}
\usepackage{gensymb}
\usepackage{rotating}
\usepackage{threeparttable}
\usepackage{comment}
\usepackage[normalem]{ulem}
\usepackage[hidelinks]{hyperref} % allows hyperlinking for references
\usepackage[version=3]{mhchem} % Formula subscripts using \ce{}
\usepackage{xcolor}
\usepackage{amsmath,amssymb,amsthm}
\usepackage{mathtools,physics}
\usepackage{subcaption}
\usepackage{tikz}

\usetikzlibrary{positioning, fit, shapes, shapes.geometric}
\usepackage{lineno}
\usepackage{xr-hyper}
\usepackage{xr}
\makeatletter

\tikzstyle{process} = [rectangle, rounded corners, minimum width=4cm, minimum height=1cm, text centered, draw=black, font=\small]
\tikzstyle{data} = [rectangle, dashed, rounded corners, minimum width=4cm, minimum height=1cm, text centered, draw=black, font=\small]
\tikzstyle{parallelogram} = [trapezium, trapezium left angle=70, trapezium right angle=110, minimum width=3.5cm, minimum height=1cm, draw=black, text centered, font=\small]
\tikzstyle{arrow} = [thick,->,>=stealth]

\newcommand*{\addFileDependency}[1]{% argument=file name and extension
\typeout{(#1)}% latexmk will find this if $recorder=0
\@addtofilelist{#1}
\IfFileExists{#1}{}{\typeout{No file #1.}}
}\makeatother

\newcommand*{\myexternaldocument}[1]{%
\externaldocument{#1}%
\addFileDependency{#1.tex}%
\addFileDependency{#1.aux}%
}
\myexternaldocument{SI}

\title{Analyzing Initialization Strategies for the Local Unitary Cluster Jastrow Ansatz within the Quantum-Centric Supercomputing Framework}

\author{Grier M. Jones}
\affiliation[UTSG ECE]{
The Edward S. Rogers Sr. Department of Electrical and Computer Engineering, 
University of Toronto, 
10 King's College Road, Toronto, Ontario, 
Canada M5S 3G4}
\alsoaffiliation[UTM CHEM]{
Department of Chemical and Physical Sciences, 
University of Toronto Mississauga, 
3359 Mississauga Road, Mississauga, Ontario, 
Canada L5L 1C6}
\email{grier.jones@utoronto.ca}

\author{Maforikan J. Amoussou}
\affiliation[UTSG ECE]{
The Edward S. Rogers Sr. Department of Electrical and Computer Engineering, 
University of Toronto, 
10 King's College Road, Toronto, Ontario, 
Canada M5S 3G4}
\alsoaffiliation[contribution]{These authors contributed equally}

\author{Maximilian O. Leach}
\affiliation[UTSG ECE]{
The Edward S. Rogers Sr. Department of Electrical and Computer Engineering, 
University of Toronto, 
10 King's College Road, Toronto, Ontario, 
Canada M5S 3G4}
\alsoaffiliation[contribution]{These authors contributed equally}

\author{Hans-Arno~Jacobsen}
\affiliation[UTSG ECE]{
The Edward S. Rogers Sr. Department of Electrical and Computer Engineering, 
University of Toronto, 
10 King's College Road, Toronto, Ontario, 
Canada M5S 3G4}
\alsoaffiliation[UTSG CS]{
Department of Computer Science, 
University of Toronto, 
40 St. George Street, Toronto, Ontario, 
Canada M5S 2E4}
\email{jacobsen@eecg.toronto.edu}

\begin{document}

\begin{tocentry}

\begin{figure}[H]
    \centering
    \includegraphics[width=\linewidth]{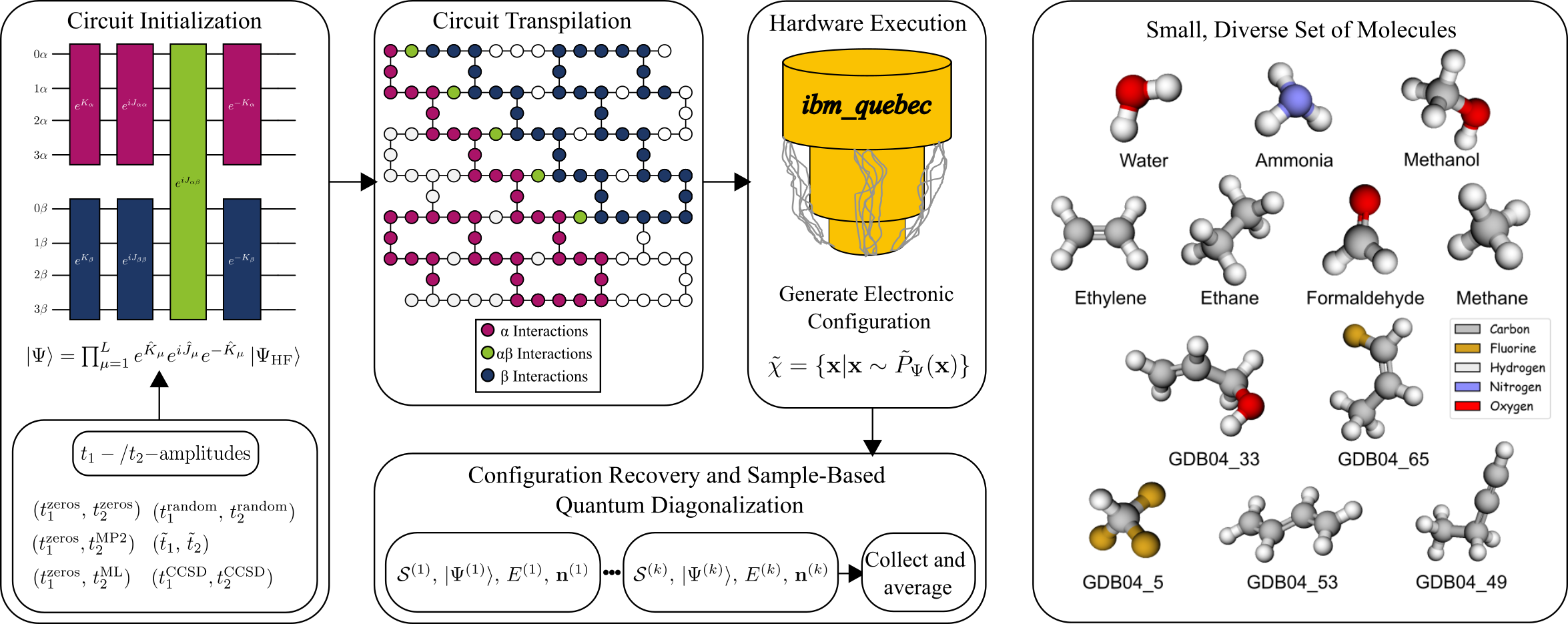}
    \caption{}
    \label{fig:TOC}
\end{figure}

\end{tocentry}

\begin{abstract}
In this study, we analyze the choice of local unitary cluster Jastrow (LUCJ) ansatz initialization and sensitivity of the sample-based quantum diagonalization (SQD) algorithm within the quantum-centric supercomputing (QCSC) framework.
We examine six initialization strategies, including those based on coupled-cluster singles and doubles (CCSD), Møller-Plesset second-order perturbation theory (MP2), data-driven coupled-cluster (DDCC), and trivial (zeroes and random) initializations, across twelve molecular systems and three basis sets (STO-3G, cc-pVDZ, and aug-cc-pVDZ).
% We find that while the mean absolute percentage errors (MAPEs) between the alternative and CCSD-initialized $t_{2}$-amplitudes span many orders of magnitude, the resulting SQD energies are largely insensitive to this variation, with most initializations recovering energies within chemical accuracy ($\pm1.6$ m$E_{\mathrm{h}}$) of the CCSD reference, particularly for larger basis sets. 
We find that while the mean absolute percentage errors (MAPEs) between the alternative and CCSD-initialized 
$t_{2}$-amplitudes span many orders of magnitude, the resulting SQD energies are largely insensitive to this variation. In particular, most initializations recover energies within chemical accuracy ($\pm1.6$ m$E_{\mathrm{h}}$) of the CCSD reference, with convergence improving as the basis set size increases.
Notably, random initialization achieves performance competitive with CCSD across all basis sets, while zeroes initialization, despite having smaller deviations from CCSD, yields the worst energy agreement.
Our results highlight that the proximity to the CCSD initialization is not a reliable predictor of the quality of electronic energies. 
These findings establish that configuration recovery within SQD, rather than circuit initialization, is the dominant factor governing energy accuracy, and suggest that computationally cheaper initialization strategies are viable alternatives to CCSD for QCSC workflows.
\end{abstract}


%GMJ Suggested introduction outline
%\begin{itemize}
%    \item \sout{General introduction (aimed at quantum computing for electronic structure)}
%    \item \sout{LUCJ and related work}
%    \item \sout{Data-driven/ML directly related}
%    \item \sout{What we introduce}
%\end{itemize}

% LUCJ circuits explanations (amplitudes relationship with angles on rotation gates, etc.)
%SQD Portion of Theory
%SQD Computational Details under "Electronic Structure Calculations"
%Methods Revision

\section{Introduction}
In molecular electronic structure calculations, a common task is to find the ground-state energy using the Schr\"{o}dinger equation with a nonrelativistic Hamiltonian in the absence of an external field, invoking the Born-Oppenheimer approximation.
In principle, this can be solved exactly using full configuration interaction (FCI) within a finite basis set; however, in practice, FCI is computationally intractable for systems with active spaces (denoted as the number of electrons $e$ in the number of spatial orbitals $o$) beyond $(22e,22o)$~\cite{vogiatzis_pushing_2017} and $(26e,23o)$~\cite{gao_distributed_2024} due to the exponential scaling of the FCI expansion.
This limitation necessitates approximate methods, such as the density matrix renormalization group~\cite{baiardi2020density,schollwock2005density}, variational two-electron reduced density matrix methods~\cite{eugene2024variational}, and Monte Carlo methods~\cite{austin2012quantum}, among others~\cite{cao_quantum_2019}.

Due to the inherent quantum nature of molecular electronic structure theory, quantum chemistry is often described as a natural application of quantum computing~\cite{reiher2017elucidating}, where quantum algorithms, such as quantum phase estimation (QPE)~\cite{aspuru-guzik_simulated_2005,abrams_simulation_1997,abrams_quantum_1999,lanyon_towards_2010,whitfield_simulation_2011,aspuru-guzik_photonic_2012} and the variational quantum eigensolver (VQE)~\cite{peruzzo_variational_2014,cerezo_variational_2021,mcclean_theory_2016,bharti_noisy_2022}, are proposed to find the ground-state energy.
While both approaches offer several advantages, such as QPE offering a suspected exponential speedup over classical methods and VQE offering a hybrid quantum-classical approach that can run on noisy near-term devices, both are ultimately constrained by the number and quality of qubits available on current hardware.
To bypass some of the computational constraints of VQE, Kanno \textit{et al.}~\cite{kanno_quantum-selected_2023,nakagawa2024adapt} 
propose the quantum selected configuration interaction (QSCI) method to leverage the quantum device for determinant selection to reduce the quantum workload.
More recently, Robledo-Moreno \textit{et al.}~\cite{robledo-moreno_chemistry_2025} introduce sample-based quantum diagonalization (SQD), which integrates a quantum-centric supercomputing (QCSC) framework~\cite{alexeev_quantum-centric_2024} that combines quantum hardware with classical supercomputing.
In their work, they demonstrate a practical, hardware-efficient approach to the molecular electronic structure problem on near-term quantum devices by leveraging the local unitary cluster Jastrow (LUCJ) ansatz~\cite{D3SC02516K}.

While the SQD approach applies to a variety of systems, including potential energy curves of the nitrogen molecule and model iron-sulfur clusters~\cite{robledo-moreno_chemistry_2025}, spin state energetics~\cite{nutzel2025solving,liepuoniute2025quantum}, excited states~\cite{barison2025quantum}, and systems using implicit solvation~\cite{kaliakin2025implicit}, possible limitations can arise that depend on the initial state preparation of the LUCJ ansatz.
One such limitation, and the main focus of this article, is the standard initialization of the LUCJ ansatz based on $t_{1}$- and $t_{2}$-amplitudes from coupled-cluster singles and doubles (CCSD).
This represents a potential bottleneck as system size increases since the method scales as $\mathcal{O}(N^6)$, where $N$ denotes the system size.
In practice, the main bottleneck of solving the CC amplitude equations is finding a solution for the $t_{2}$-amplitudes, where the computational overhead is due to each molecule containing $N_{\mathrm{occ}}^2 N_{\mathrm{virt}}^2$ $t_{2}$-amplitudes, where $N_{\mathrm{occ}}$ and $N_{\mathrm{virt}}$ denote the number of occupied and virtual orbitals, respectively.
To this end, methods based on machine learning, such as data-driven coupled-cluster (DDCC)~\cite{doi:10.1021/acs.jpclett.9b01442,doi:10.1021/acs.jpca.4c05718,JONES2023509}, have been proposed to reduce the computational overhead of predicting $t_{2}$-amplitudes.

While the previously mentioned limitation concerns initial state preparation, a second potential limitation concerns the robustness of the configuration recovery scheme in SQD to noise.
As highlighted by Vaquero-Sabater \textit{et al.}~\cite{vaquero2026noise}, results within the QSCI framework indicate that sampling, noise, and configuration recovery are more critical than the choice of ansatz.
Their work highlights the limitations of LUCJ and SQD methods to near-term, noisy devices, since noiseless sampling of the LUCJ ansatz can produce highly compact and biased configurational spaces.
We consider this dynamic interplay between hardware noise and classical post-processing a fundamental factor governing the accuracy and structural resilience of SQD, with configuration recovery being the dominant factor in recovering accurate electronic energies.

In this study, we show the weak dependence of state preparation on model performance by analyzing six different initialization strategies, including the standard CCSD implementation, one based on M{\o}ller-Plesset second-order perturbation theory (MP2), and two based on DDCC, along with zeroes and random initialization.
% Our work highlights that the quality of energy obtained by SQD depends more strongly on system and basis set size than on the choice of circuit initialization.
Our results indicate that SQD energy quality is robust to circuit initialization choices, with performance primarily determined by system and basis set size.
% Throughout this work, we compare the standard CCSD initialization with five alternative initializations and all six initialization strategies against classical methods, such as complete active space configuration interaction (CASCI) and heat-bath configuration interaction (HCI)~\cite{holmes_heat-bath_2016}.
Throughout this work, we compare the standard CCSD initialization against five alternative strategies, and evaluate all six against classical methods such as complete active space configuration interaction (CASCI)~\cite{Roos2007-ko} and heat-bath configuration interaction (HCI)~\cite{holmes_heat-bath_2016}.
Additionally, we examine seven small molecules with active spaces that admit exact diagonalization and five molecules where CASCI is intractable.
Our findings establish that the proximity of alternative initializations to the CCSD initialization does not strictly dictate variational energy accuracy, revealing that SQD deployed within the QCSC framework is weakly dependent on the choice of ansatz initialization.

\section{Theoretical aspects}
\subsection{Electronic Structure Problem}\label{subsection:electronicstructure}
To obtain the ground state energy in molecular electronic structure calculations, the Schr\"{o}dinger equation is solved using the non-relativistic Born-Oppenheimer Hamiltonian in the absence of an external field, defined as
\begin{equation}
    \Hat{H} = \sum_{\substack{pr \\ \sigma}} h_{pr} \Hat{a}^{\dagger}_{p\sigma}\Hat{a}_{r\sigma} + \frac{1}{2} \sum_{\substack{prqs \\ \sigma\tau}} (pr\vert qs) \Hat{a}^{\dagger}_{p\sigma}\Hat{a}^{\dagger}_{q\tau}\Hat{a}_{s\tau}\Hat{a}_{r\sigma}.
    \label{eq:spinfreeHamiltonian}
\end{equation}
where $p,q,r,s$ denote general spatial-orbitals, $\sigma,\tau$ denote spin-functions, $h_{pq}$ and $(pr\vert qs)$ are one- and two-electron integrals, and $\Hat{a}^{\dagger}_{p\sigma}$ and $\Hat{a}_{p\sigma}$ denote Fermionic creation and annihilation operators acting on spin-orbital $p\sigma$.
Similarly, in second quantization, the wave function is represented by an occupation-number (ON) vector defined as
\begin{equation}
    \ket{\Psi} = \prod_{p\sigma} (\Hat{a}^{\dagger}_{p\sigma})^{x_{p\sigma}} \ket{\mathrm{vac}},
    \label{eq:ONVector}
\end{equation}
where $\ket{\mathrm{vac}}$ is the vacuum state, i.e., a wave function containing zero electrons, and $x_{p\sigma} \in \{0,1\}$.
For example, a restricted Hartree-Fock (RHF) wave function can be defined as
\begin{equation}
    \ket{\Psi_{\mathrm{RHF}}} = \ket{\underbrace{0\ldots0}_{N_{\mathrm{MO}-N_{\beta}}} \underbrace{1\ldots1}_{N_{\beta}} \underbrace{0\ldots0}_{N_{\mathrm{MO}}-N_{\alpha}} \underbrace{1\ldots1}_{N_{\alpha}}},
    \label{eq:RHFExample}
\end{equation}
where the wave function is partitioned into two spin sectors, one for the $\alpha$ ($m_s=+\frac{1}{2}$) and one for the $\beta$ ($m_s=-\frac{1}{2}$) spins.
To map ON vector onto an element of the computational basis, we use the Jordan-Wigner (JW) Fermion to qubit mapping, where $M$ spin-orbitals are mapped to $M$-qubits~\cite{jordan1928paulische}.
Within the SQD algorithm, the occupation number is an important property of the bitstrings since it is required for configuration recovery, where for a given electronic configuration $\Psi$, the number of spin-$\sigma$ electrons is defined a $N_{\Psi\sigma} = \sum_{p} \Psi_{p\sigma}$ and the total number of electrons is defined as $N_{\Psi}=\sum_{\sigma}N_{\Psi\sigma}$.
In computer science parlance, $N_{\Psi}$ is the \textit{Hamming weight} of the bitstring $\Psi$.

\subsection{Ansatz and Initialization Strategies}
Like in the study of Robledo-Moreno \textit{et al.}~\cite{robledo-moreno_chemistry_2025}, we use the LUCJ ansatz, a hardware-efficient and physically intuitive wave function for solving chemically relevant problems~\cite{motta_bridging_2023}.
In this section, we will start our discussion by reviewing the unitary cluster Jastrow (UCJ) ansatz and by extension, the LUCJ ansatz.
Next, we will highlight the relationship between the LUCJ ansatz and traditional coupled-cluster, which is often used to initialize the LUCJ ansatz in QCSC frameworks.
This relationship will then be used to motivate the introduction of the data-driven LUCJ ansatz, which incorporates the machine-learning-based DDCC method\cite{doi:10.1021/acs.jpclett.9b01442}, to create a more efficient initialization of the LUCJ circuit parameters, along with other alternative initializations examined in this study.

While our work focuses on a QCSC framework, the LUCJ ansatz was originally proposed as a variational ansatz analogous to a restricted HF state time-evolved under a Hubbard Hamiltonian.
This approach combines hardware efficiency with physical intuition to provide an ansatz that can be adapted to various device qubit topologies.\cite{motta_bridging_2023}
The LUCJ ansatz applies local constraints to a UCJ wave function; the UCJ formulation incorporates coupled-cluster theory with Jastrow factors, i.e., symmetric scalar functions of electron positions used to encode electron correlation.\cite{foulkes2001quantum}
As with many post-Hartree-Fock approaches, the zeroth-order wave function used to initialize the LUCJ ansatz is a Hartree-Fock (HF) reference ($\ket{\Psi_{\mathrm{HF}}}$), where the UCJ wave function is formulated as a product of $L$ layers applied to a HF reference, defined as 
\begin{equation}
    \ket{\Psi} = \prod^L_{\mu=1} e^{\hat{K}_{\mu}} e^{i\hat{J}_\mu} e^{-\hat{K}_{\mu}}\ket{\Psi_{\mathrm{HF}}},
    \label{eqn:lucj}
\end{equation}
where $\hat{K}_{\mu}$ is a unitary orbital rotation operator,
\begin{equation}
    \hat{K}_\mu = \sum_{pq,\sigma}{}{{K_{pq}}^\mu}\hat{a}^\dagger_{p\sigma}\hat{a}_{q\sigma}
    \label{eq:KHat}
\end{equation}
and $\hat{J}_{\mu}$ is a Jastrow factor operator, 
\begin{equation}
    \hat{J}_\mu = \sum_{pq,\sigma\tau}{{J_{pq,\sigma\tau}}^\mu}\hat{n}^\dagger_{p\sigma}\hat{n}^\dagger_{q\tau}.
    \label{eq:JHat}
\end{equation}
In Equations \ref{eq:KHat} and \ref{eq:JHat}, the indices $p$ and $q$ denote spatial molecular orbitals, $\sigma$ and $\tau$ denote electron spin.
The matrix element ${K_{pq}}^\mu$ belongs to a unitary one-particle rotation matrix, $\mathbf{K}$, and ${J_{pq,\sigma\tau}}^\mu$ are matrix elements of the matrix $\mathbf{J}$ corresponding to Jastrow factor weights.
In the LUCJ ansatz, local approximations are incorporated through modifications of the opposite- and same-spin number-number operators in the $\hat{J}_\mu$ operator (Eq. \ref{eq:JHat}).
This local approximation enables the ansatz to be tailored to specific device topologies, including linear, square, hexagonal, and heavy-hex topologies.
For more details regarding the original implementation of the LUCJ ansatz, refer to \citet{motta_bridging_2023}

One key aspect of the LUCJ ansatz is that the circuit parameters can be initialized using truncated doubly factorized low-rank decomposed $t_{2}$-amplitudes from MP2 or CCSD.\cite{doi:10.1021/acs.jctc.9b00963,10.1063/1.4829536,doi:10.1021/acs.jctc.6b00288}
Using either MP2, which scales as $\mathcal{O}(N^5)$, or CCSD, which scales as $\mathcal{O}(N^6)$, can be viewed as a bottleneck for initializing the LUCJ ansatz as system size increases.
To this end, we will provide a brief overview of the DDCC method and how it can be incorporated to reduce the cost of initializing the LUCJ ansatz.
We start this overview by defining the coupled-cluster wave function,
\begin{align}
    \ket{\Psi_{\mathrm{CC}}} = \exp(\hat{T}) \ket{\Psi_{\mathrm{HF}}},
\label{CC ansatz}
\end{align}
where $\hat{T}$ is the cluster operator, and $\ket{\Psi_{\mathrm{HF}}}$ is a HF reference state.
To reduce computational cost, the cluster operator is often truncated, where the CCSD cluster operator is truncated to include single ($\hat{T}_{1}$) and double ($\hat{T}_{2}$) excitations:
\begin{equation}
    \hat{T} = \hat{T}_{1} + \hat{T}_{2} = \sum_{ia} t_{i}^{a} \hat{a}^{\dagger}_{a} \hat{a}_{i} + \frac{1}{4} \sum_{ijab} t_{ij}^{ab} \hat{a}^{\dagger}_{a}\hat{a}_{i}\hat{a}^{\dagger}_{b} \hat{a}_{j}.
    \label{eq:CCSDclusterop}
\end{equation}
In Equation \ref{eq:CCSDclusterop}, $\hat{a}^{\dagger}_{a}$ and $\hat{a}^{\dagger}_{b}$ are Fermionic creation operators, $\hat{a}_{i}$ and $\hat{a}_{j}$ are Fermionic annihilation operators, $i$ and $j$ denote occupied orbitals, $a$ and $b$ denote virtual orbitals.
Additionally, $t_{i}^{a}$ are referred to as $t_{1}$-amplitudes and are often initialized as zeroes, while $t_{ij}^{ab}$ are referred to as $t_{2}$-amplitudes and are initialized using MP2 $t_{2}$-amplitudes, defined as 
\begin{equation}
    {t_{ij}^{ab}}_{\mathrm{(MP2)}} = \frac{\Braket{ij || ab }}{(\varepsilon_i + \varepsilon_j - \varepsilon_a - \varepsilon_b)}.
\label{MP2 t2}
\end{equation}

In a conventional coupled-cluster scheme, the set of amplitudes in the cluster operator is solved iteratively using the projected coupled-cluster equations.
As system size increases, the cost of solving the projected coupled-cluster equations becomes prohibitive; the DDCC method has therefore been proposed as a machine-learning-based alternative that can either bypass CCSD optimization entirely (as in our ML initialization) or provide a warm-start solution that converges in fewer iterations than MP2 (as in our ML{\_}exact initialization).
The feature set consists of the MP2 $t_{2}$-amplitudes, along with the numerator, or antisymmetrized two-electron integrals ($\Braket{ij || ab }$), denominator, or deviation between the occupied and virtual orbital energies ($\varepsilon_i + \varepsilon_j - \varepsilon_a - \varepsilon_b$), and with their individual contributions $\varepsilon_{i}$, $\varepsilon_{j}$, $\varepsilon_{a}$, $\varepsilon_{b}$.
Additionally, the broken-down contributions to the orbital energies, including the matrix elements of the one-electron Hamiltonian ($h$), Coulomb ($J$), and exchange ($K$) matrices, are included in the feature set.
Other features include a binary feature to denote whether two electrons go to the same virtual indices (i.e., $a=b$) and the Coulomb and exchange integrals ($J_{i}^{a}$, $J_{j}^{b}$, $K_{i}^{a}$, $K_{j}^{b}$), for a total of 30 features that encode the electronic structure properties of a molecule.

Unlike the original DDCC scheme, we highlight a subset of the five most important features, as determined by SHapley Additive exPlanations (SHAP) values.\cite{9265985}
SHAP analysis uses cooperative game theory to determine feature importance, where the SHAP values from the top five features include the initial MP2 $t_{2}$-amplitude (${t_{ij}^{ab}}_{\text{(MP2)}}$), the logarithmic magnitude of the MP2 $t_{2}$-amplitude ($\log (| {t_{ij}^{ab}}_{\text{(MP2)}}|)$), the numerator ($\Braket{ij || ab }$) and denominator terms ($\varepsilon_i + \varepsilon_j - \varepsilon_a - \varepsilon_b$) in Equation \ref{MP2 t2} , and the binary feature ($\delta_{ab}$) that describes whether the excited electrons occupy the same virtual orbital.
For details on the SHAP analysis, see Supporting Information (SI) Section \ref{section:feature_analysis}.

In this study, we examine six strategies to initialize the $t_{2}$-amplitudes.
The first is how CCSD is initialized, that is, using MP2 $t_{2}$-amplitudes, where the set of amplitudes is defined as $t_{2}^{\text{MP2}}$.
The second is how the LUCJ ansatz is initialized in \citet{robledo-moreno_chemistry_2025}, where the $t_{2}$-amplitudes are produced by CCSD, denoted as $t_{2}^{\text{CCSD}}$.
The third initialization method incorporates the DDCC scheme into the LUCJ workflow, where the ML-produced $t_{2}$-amplitudes are directly incorporated into the LUCJ quantum circuit.
The fourth approach combines the second and third approaches, where the ML-produced $t_{2}$-amplitudes are injected into the coupled-cluster iterative solver and the optimized set of the $t_{2}$-amplitudes are denoted as $\Tilde{t}_{2}$, and are numerically equivalent to $t_{2}^{\text{CCSD}}$-amplitudes up to a tolerance.
The fifth and sixth cases are extremes to understand the effects of initialization on the QCSC workflow, where the $t_{2}$-amplitudes that are initialized using zeroes are denoted as $t_{2}^{\text{zeroes}}$ and randomly initialized $t_{2}$-amplitudes are denoted as $t_{2}^{\text{random}}$.
As with the $t_{2}$-amplitudes,  we explore four initializations for the $t_{1}$-amplitudes.
% Like the $t_{2}$-amplitudes, we explore four unique initializations for the $t_{1}$-amplitudes.
The first initialization we examine is $t_{1}$-amplitudes set to zero ($t_{1}^{\text{zeroes}}$), as is done at the start of CCSD calculations.
The second set correspond to $t_{1}$-amplitudes produced after the optimization of the coupled-cluster equations ($t_{1}^{\text{CCSD}}$).
The third initialization corresponds to the set of optimized $t_{1}$-amplitudes produced by solving the coupled-cluster equations using the injected $t_{2}$-amplitudes ($\Tilde{t}_{1}$; which are in principle equivalent to $t_{1}^{\text{CCSD}}$).
The final $t_{1}$-amplitudes is an edge case---a random initialization, denoted $t_{1}^{\text{random}}$---included to examine the sensitivity of the workflow to the initial parameter set.
% The last initialization we explore for the $t_{1}$-amplitudes is an edge case to examine the effects of the initial parameter set on the overarching workflow and corresponds to a random initialization, denoted as $t_{1}^{\text{random}}$.
In total, six unique combinations are explored in this study: $(t_{1}^{\text{zeroes}}, t_{2}^{\text{MP2}})$, $(t_{1}^{\text{CCSD}}, t_{2}^{\text{CCSD}})$, $(t_{1}^{\text{zeroes}}$, $t_{2}^{\text{ML}})$, $(\Tilde{t}_{1}$, $\Tilde{t}_{2})$, $(t_{1}^{\text{zeroes}}$, $t_{2}^{\text{zeroes}})$, and $(t_{1}^{\text{random}}$, $t_{2}^{\text{random}})$, referred to as the MP2, CCSD, ML, ML\_exact, zeroes, and random initialization, respectively, throughout this work.

\subsection{Sample-Based Quantum Diagonalization}
\begin{figure}[htbp]
    \centering
    \includegraphics[width=\linewidth]{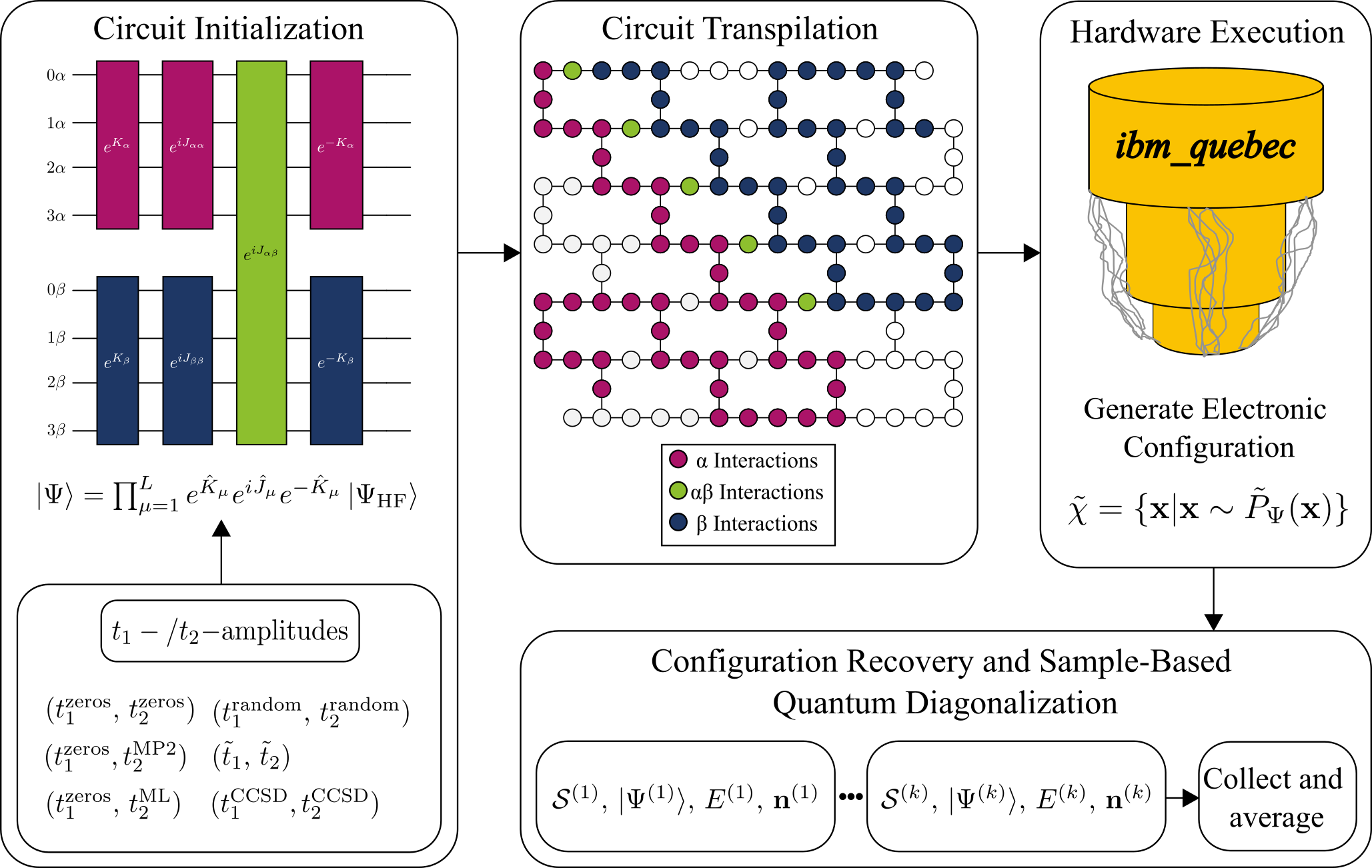}
    \caption{Schematic diagram of the proposed workflow to analyze circuit initialization within the QCSC framework.}
    \label{fig:workflow}
\end{figure}

After choosing an initialization scheme for the $t_{1}$- and $t_{2}$-amplitudes, the remaining steps of the algorithm follow those of~\citet{robledo-moreno_chemistry_2025} (Figure~\ref{fig:workflow}), where the LUCJ circuit is run on a prefault-tolerant quantum computer to generate a set of electronic configurations.
After mapping the quantum circuit to hardware, the next steps include hardware execution and SQD. 
Due to noise on prefault-tolerant quantum computers, the wave function $\ket{\Psi}$ (Eq. \ref{eqn:lucj}) is prepared multiple times and measured in the computational basis to form a set of noisy measurement configurations, defined as
\begin{equation}
    \Tilde{\chi} = \{ \Psi \vert \Psi \sim \Tilde{P}_{\Psi}(\Psi) \},
    \label{eq:NoisyDistribution}
\end{equation}
\noindent where the bitstrings represent electronic configurations denoted as $\Psi\in \{0,1\}^{N}$ and $\Tilde{P}_{\Psi}(\Psi)=\mel{\Psi}{\Tilde{\rho}}{\Psi}$, such that $\Tilde{\rho}=\ket{\Tilde{\Psi}}\bra{\Tilde{\Psi}}$, i.e., a noisy density matrix.
Similarly, a set of noiseless configurations is defined as 
\begin{equation}
    \chi = \{ \Psi \vert \Psi \sim P_{\Psi}(\Psi) \},
    \label{eq:NoiselessDistribution}
\end{equation}
\noindent where $P_{\Psi}(\Psi) = \vert \braket{\Psi}{\Psi} \vert^{2}$.
The distribution $P_{\Psi}$ will include \textit{deadwood configurations} that do not contribute to the desired low-energy states due to noise in the quantum system~\cite{ivanic2001identification}.
To this end, the self-consistent configuration recovery process has been introduced to perform a partial probabilistic recovery of noiseless configurations from $\Tilde{\chi}$.
Configuration recovery targets configurations $\Psi$ that contain the wrong particle number, i.e., $N_{\Psi}$ does not match the desired particle number specified by the charge and spin-state of the system, and corrects them through an iterative process to generate a set of recovered configurations $\chi_{\mathrm{R}}$ used in the following subspace diagonalization step.
At each step of the self-consistent configuration recovery process, the orbital occupancies for each spin-orbital ($p\sigma$) are obtained by averaging over the $K$ batches:
\begin{equation}
    n_{p\sigma} = \frac{1}{K}\sum_{1\le k\le K} \mel{\Psi^{(k)}}{\Hat{n}_{p\sigma}}{\Psi^{(k)}}.
\end{equation}
Lastly, we note that the initial set of orbital occupations, denoted as $\mathbf{n}$ for the full set of orbitals, is obtained using the raw quantum samples with the correct number of particles, or Hamming weights. 

Following the configuration recovery step, $K$ batches of $d$ configurations $\mathcal{S}^{(1)},\ldots,\mathcal{S}^{(K)}$ are drawn from a distribution proportional to the empirical frequencies of each $\Psi$ in $\chi_{R}$.
This step draws inspiration from classical~\cite{evangelista2014adaptive,holmes_heat-bath_2016,holmes2016efficient,tubman2016deterministic,schriber2016communication,schriber2017adaptive,sharma_semistochastic_2017} and quantum selected configuration interaction (SCI)~\cite{kanno_quantum-selected_2023,nakagawa2024adapt} approaches, where the Hamiltonian is projected and diagonalized over each $\mathcal{S}^{(k)}: k=1,\ldots, K$.
The projected many-body Hamiltonian is defined as
\begin{equation}
    \Hat{H}_{\mathcal{S}^{(k)}} = \Hat{P}_{\mathcal{S}^{(k)}} \Hat{H} \Hat{P}_{\mathcal{S}^{(k)}}
    \label{ed:projectedHamiltonian}
\end{equation}
where the projection operator is defined as $\Hat{P}_{\mathcal{S}^{(K)}}=\sum_{\Psi\in \mathcal{S}^{(K)}} \ket{\Psi}\bra{\Psi}$.
For each batch, the corresponding energies and wave function are labeled $E^{(k)}$ and $\ket{\Psi^{(k)}}$, respectively, and solved using the traditional Davidson algorithm~\cite{davidsorq1975theiterative}.

\section{Computational Details}
% Computational Details Checklist
% \begin{itemize}
%     \item Classical electronic structure
%     \begin{itemize}
%         \item Molecules (spins, charge, names, etc.)
%         \item Active spaces
%         \item Basis sets
%         \item What packages generated what
%         \begin{itemize}
%             \item Psi4/Psi4NumPy
%             \item PySCF
%         \end{itemize}
%     \end{itemize}
%     \item Machine Learning
%     \begin{itemize}
%         \item Scikit-Learn (feature scaling and model)
%         \item SHAP (feature reduction)
%     \end{itemize}
%     \item Quantum
%     \begin{itemize}
%         \item \textsc{ffsim}\cite{ffsim}: LUCJ 
%         \item SQD (need parameters)  Method citations\cite{kanno_quantum-selected_2023,nakagawa2024adapt,robledo-moreno_chemistry_2025}, code citation\cite{saki2024qiskit}
%         \item device names (format \textit{ibm{\textunderscore}name})
%         \item Runtime parameters (optimization and resilience levels + shots)
%     \end{itemize}
% \end{itemize}

\subsection{Electronic Structure Theory Calculations}
Our workflow incorporates two established electronic structure packages \textsc{Psi4}\cite{smith2020psi4} and \textsc{PySCF}\cite{sun_pyscf_2018,sun2020recent}.
\textsc{Psi4} and \textsc{Psi4NumPy}\cite{Smith2018-vo} are used to generate all electronic structure data required to integrate the DDCC method into the LUCJ workflow, including HF, MP2, and spin-factored CCSD calculations. 
All reference CASCI~\cite{Roos2007-ko} calculations were performed using \textsc{PySCF} version 2.10.0~\cite{sun_pyscf_2018, sun2020recent}, while reference HCI calculations were performed using \textsc{Dice}~\cite{holmes_heat-bath_2016, sharma_semistochastic_2017} via the Dice Eigensolver Qiskit-addon (\href{https://qiskit.github.io/qiskit-addon-dice-solver/}{https://qiskit.github.io/qiskit-addon-dice-solver/}).
Throughout this work, we use the STO-3G~\cite{hehre_self-consistent_1969, collins_self-consistent_1976}, cc-pVDZ~\cite{dunning1989a, prascher2011a, woon1994a}, and aug-cc-pVDZ\cite{1992JChPh..96.6796K} basis sets.

We initialize the LUCJ ansatz parameters by mapping the amplitudes onto the one-body ($\Hat{K}^{\mu}$) and density--density ($\Hat{J}^{\mu}$) operators in Eq.~\ref{eqn:lucj} with the number of ansatz layers varied from $L=1, \ldots,5$.
All initialization calculations were performed using \textsc{PySCF} and circuit construction was carried out using \textsc{ffsim} version 0.0.60~\cite{ffsim,sung2026ffsim}.
Quantum circuits were constructed and transpiled using an optimization level of 3, which is the highest level of circuit optimization, and a resilience level of 0, or no error mitigation, in \textsc{Qiskit} version 2.2.1~\cite{qiskit2024}.
All calculations were executed using 10,000 shots (i.e., individual circuit executions) on the \texttt{ibm\_quebec} 156-qubit Heron R2 superconducting processor. 
All SQD calculations were performed with 10 batches, 1,000 samples per batch, and 5 iterations, using the \textsc{Fulqrum}~\cite{nation2026generalized} framework to improve computational efficiency.

\subsection{Machine Learning Models}
All electronic structure features used in the DDCC models are scaled using the MinMaxScaler from SciKit-Learn~\cite{scikit-learn} and trained using the XGBoostRegressor module from the xgboost package.\cite{10.1145/2939672.2939785} 
Grid-search cross-validation was used to fine-tune the model hyperparameters, including the maximum tree depth, number of estimators, and L1 and L2 regularization.
All optimized hyperparameters are listed in Section \ref{section:ml} of the Supporting Information and all reported performance metrics correspond to the performance on the test set unless stated otherwise.
Lastly, all ML models used throughout this study use 100 molecules in the training set.

\subsection{Molecular Structures}
\begin{figure}[htbp]
    \centering
    \includegraphics[width=0.7\linewidth]{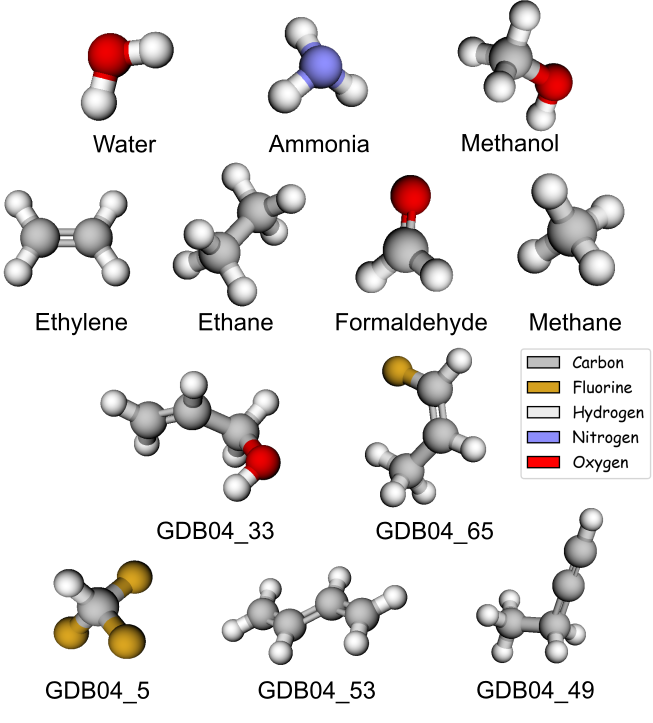}
    \caption{Molecular structures comprising the three datasets used in this study. (i) A set of smaller molecules with multiple conformers, (ii) a benchmark set containing one representative conformer per molecule excluded from training and testing, and (iii) a set of larger and more chemically diverse molecules selected from the GDB-11 database.}
    \label{fig:mols}
\end{figure}

The molecular structures used in this study can be partitioned into two sets: those used to train the DDCC model and those used in our analysis of the LUCJ initializations.
The first dataset was obtained from the original DDCC study (\href{https://gitlab.com/jtowns28/ddcc-voglab2019}{https://gitlab.com/jtowns28/ddcc-voglab2019}), where we selected 100 random water, methanol, ethylene, ethane, methane, ammonia, and formaldehyde conformers, for a total of 700 structures.~\cite{doi:10.1021/acs.jpclett.9b01442,noauthor_jacob_2019}
From the 700 conformers, 100 were randomly selected for training and 50 for testing.
The second dataset consists of one conformer per molecule from the previous dataset, ensuring that none of these were present in the training or test sets, and five larger and more complex molecules randomly selected from the GDB-11 dataset.~\cite{Fink2005-kc, Fink2007-zk} 
The molecular structures selected from the GDB-11 dataset were optimized using the MMFF94s~\cite{halgren1999mmff} force field, as implemented in ChemML.~\cite{haghighatlari2019chemml}
Examples of all the molecules examined are highlighted in Figure \ref{fig:mols}, with all active spaces, denoted as $(N_{e}, N_{o})$, where $N_{e}$ is the number of electrons and $N_{o}$ is the number of spatial orbitals, used in this study highlighted in Table \ref{tab:activespaces}.

\begin{table}[H]
    \centering
    \begin{tabular}{llllrr}
    \toprule
    molecule & formula & xyz filename & $N_e$ & $N_o$ \\
    \midrule
    ammonia & NH$_3$ & ammonia157.xyz & 8 & 10 \\
    methane & CH$_4$ & methane50.xyz & 9 & 10 \\
    ethylene & C$_2$H$_4$ & ethylene42.xyz & 14 & 16 \\
    ethane & C$_2$H$_6$ & ethane28.xyz & 16 & 18 \\
    water & H$_2$O & water183.xyz & 7 & 10 \\
    formaldehyde & CH$_2$O & formaldehyde138.xyz & 12 & 16 \\
    methanol & CH$_3$OH & methanol22.xyz & 14 & 18 \\
    fluoroform & CHF$_3$ & GDB04\_5.xyz & 21 & 34 \\
    buta-1,3-diene & C$_4$H$_6$ & GDB04\_53.xyz & 26 & 30 \\
    but-1-yne & C$_4$H$_6$ & GDB04\_49.xyz & 26 & 30 \\
     prop-2-en-1-ol & C$_3$H$_6$O & GDB04\_33.xyz & 26 & 32 \\
     (Z)-1-fluoroprop-1-ene & C$_3$H$_5$F & GDB04\_65.xyz & 25 & 32 \\
    \bottomrule
    \end{tabular}
    \caption{The active spaces examined in this study, where $(N_e, N_o)$ denotes the number of electrons, $N_e$, and spatial orbitals, $N_o$.}
    \label{tab:activespaces}
\end{table}

\section{Results}\label{section:results}
\subsection{Analysis of $t_2$-Amplitudes}\label{subsection:amplitudes}
\begin{figure}[H]
    \centering
    \begin{subfigure}{\textwidth}
        \includegraphics[width=\linewidth]{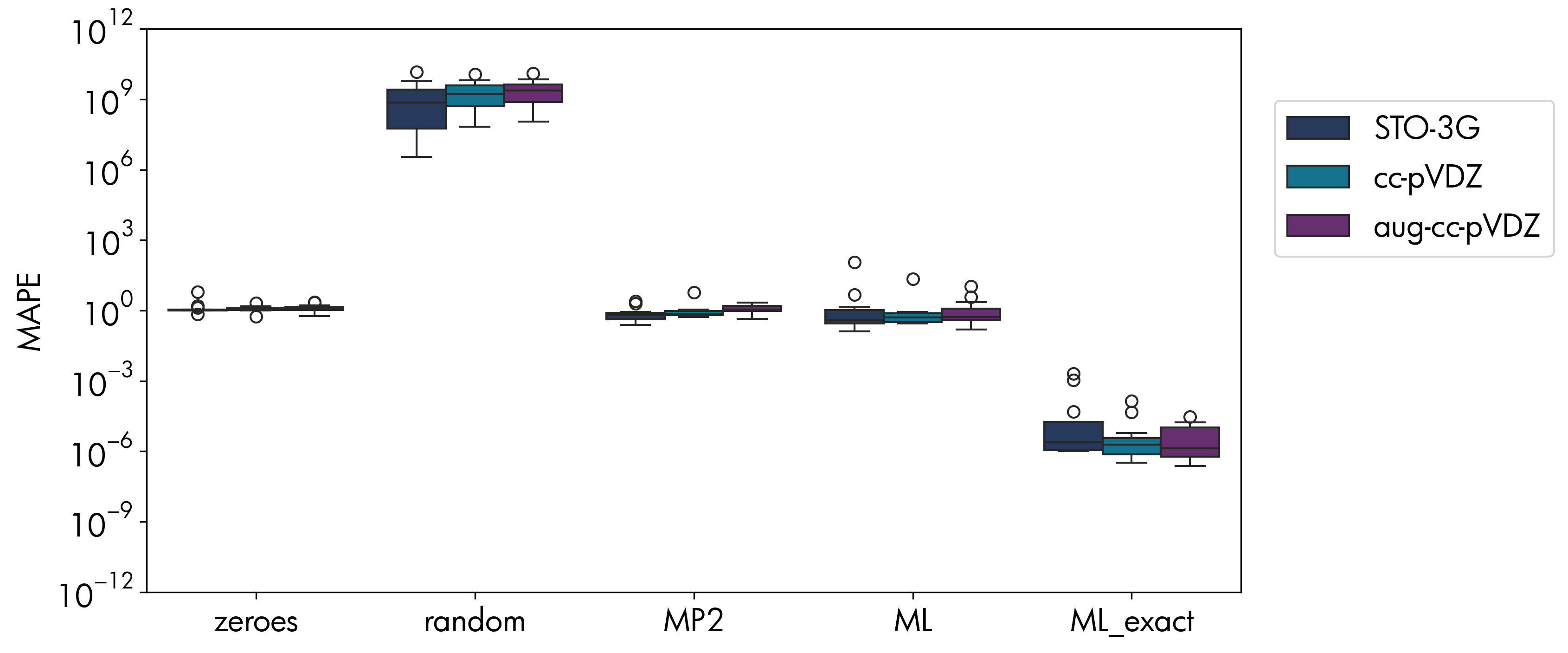}
        \caption{}
        \label{fig:AmpMapeBoxplot}
    \end{subfigure}
    \hfill
    \begin{subfigure}{\textwidth}
        \includegraphics[width=\linewidth]{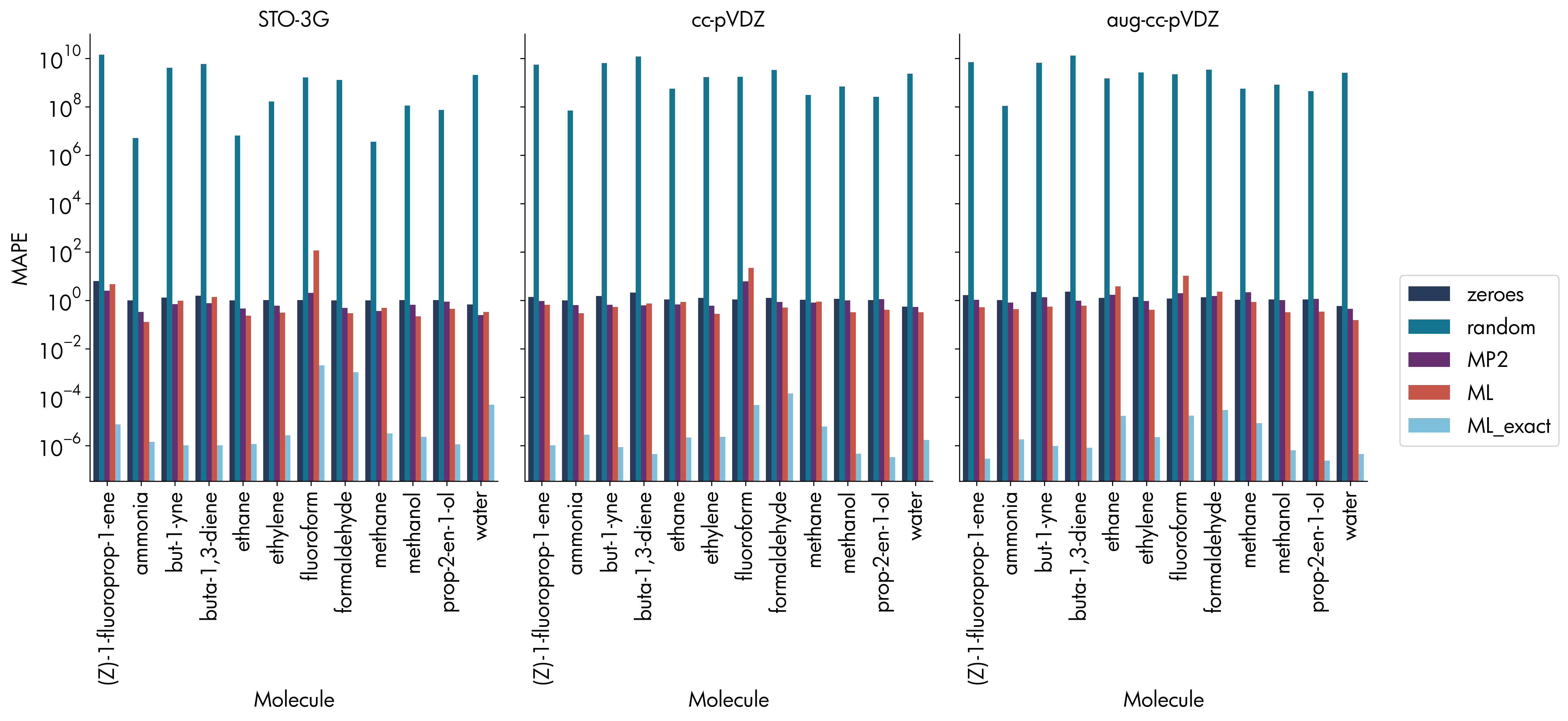}
        \caption{}
        \label{fig:AmpMapeBoxplot_b}
    \end{subfigure}
    \caption{Both figures highlight the mean absolute percent error (MAPE) between the various initialization strategies in this study and the standard CCSD initialization. (a) Shows the MAPE (y-axis) for each initialization strategy (x-axis) with respect to the basis set (hue) using a box plot. (b) Highlights the MAPE (y-axis) for each molecule (x-axis) and basis set (columns) examined using a bar plot.}
    \label{fig:AmpMape}
\end{figure}

Since the dominant cost of initializing the LUCJ ansatz is the convergence of the CCSD $t_2$-amplitudes, we start our analysis by comparing the standard initialization (CCSD) against the previously mentioned alternatives.
Additionally, since the $t_1$-amplitudes are typically initialized to zero and impose negligible computational overhead, we omit their analysis in this section.
In Figure~\ref{fig:AmpMape}, we use the mean absolute percentage error (MAPE) to highlight the deviation between the alternative ($t_{2}^{\text{zeroes}}$, $t_{2}^{\text{random}}$, $t_{2}^{\text{MP2}}$, $t_{2}^{\text{ML}}$, $\Tilde{t}_{2}$) and standard initializations ($t_{2}^{\text{CCSD}}$). 
In Figure~\ref{fig:AmpMapeBoxplot}, we plot the MAPE with respect to the choice of $t_2$-amplitudes, with the hue denoting the basis set.
% We highlight the MAPE with respect to the choice of $t_2$-amplitudes, where the hue denotes the basis set, in Figure~\ref{fig:AmpMapeBoxplot}.
Across all three basis sets, the $\Tilde{t}_{2}$-amplitudes are the closest to the $t_{2}^{\text{CCSD}}$-amplitudes due to the exact nature of the CC solver.
For these amplitudes, the MAPEs have a mean of $9.63 \times 10^{-5} \%$, a range of $2.34 \times 10^{-7}\%$ (minimum) to $2.04 \times 10^{-3}\%$ (maximum), and a standard deviation of $3.78 \times 10^{-4}\%$. 
The next closest are the $t_{2}^{\text{MP2}}$-amplitudes that are used to initialize the $t_{2}^{\text{CCSD}}$-amplitudes, where the MAPEs have a mean of $1.09\%$, a range of $2.46 \times 10^{-1}\%$ to $6.00\%$, and a standard deviation of $9.97 \times 10^{-1}\%$.
Since $t_{2}^{\text{CCSD}}$-amplitudes are typically small and centered around zero, the third closest set of $t_{2}$-amplitudes corresponds to $t_{2}^{\text{zeroes}}$, with MAPEs ranging from $5.41 \times 10^{-1} \%$ to $6.32\%$, with a mean of $1.34\%$ and standard deviation of $9.34 \times 10^{-1}\%$.
The fourth closest corresponds to $t_{2}^{\text{ML}}$, which are dependent on the training data of the ML model, where the MAPEs have a mean of $4.82\%$, range of 
$1.29 \times 10^{-1} \% $ to $ 1.16 \times 10^{2}\%$, and standard deviation of $1.94 \times 10^{1}\%$.
The initialization with the largest MAPEs corresponds to $t_{2}^{\text{random}}$, where the MAPEs have a range of $3.52\%$ to $1.44 \times 10^{10}\%$, mean of $2.91 \times 10^{9}\%$, and standard deviation of $3.72 \times 10^{9}\%$.

In Figure~\ref{fig:AmpMapeBoxplot_b}, we analyze the MAPE of each initialization with respect to the basis set and molecule. 
For the $t_{2}^{\text{zeroes}}$-amplitudes, the maximum values correspond to (Z)-1-fluoroprop-1-ene with an MAPE of $6.32\%$ for the STO-3G basis set, buta-1,3-diene with an MAPE of $2.10\%$ for the cc-pVDZ basis set, and buta-1,3-diene with an MAPE of $2.26\%$ for the aug-cc-pVDZ basis set.
The $t_{2}^{\text{random}}$-amplitudes have maximum MAPEs of $1.44 \times 10^{10}\%$ for (Z)-1-fluoroprop-1-ene using the STO-3G basis set, $1.18 \times 10^{10}\%$ for buta-1,3-diene using the cc-pVDZ basis set, and $1.29 \times 10^{10}\%$ for buta-1,3-diene using the aug-cc-pVDZ basis set.
For the STO-3G, cc-pVDZ, and aug-cc-pVDZ basis sets, the $t_{2}^{\text{MP2}}$-amplitudes have a maximum value of $2.46\%$ ((Z)-1-fluoroprop-1-ene), $6.00\%$ (fluoroform), and $2.15\%$ (methane), respectively.
Across all three basis sets, fluoroform has the largest MAPE for the $t_{2}^{\text{ML}}$-amplitudes, where the maximum MAPEs for the STO-3G, cc-pVDZ, and aug-cc-pVDZ basis sets are $1.16 \times 10^{2}\%$, $2.18 \times 10^{1}\%$, and $1.05 \times 10^{1}\%$, respectively.
Due to the nature of CC solver, the $\Tilde{t}_{2}$-amplitudes have the smallest MAPEs, where the STO-3G, cc-pVDZ, and aug-cc-pVDZ basis sets have maximum MAPEs of $2.04 \times 10^{-3}\%$ (fluoroform), $1.42 \times 10^{-4}\%$ (formaldehyde), and $2.90 \times 10^{-5}\%$ (formaldehyde), respectively.
Notably, as we discuss in Section~\ref{subsection:energies}, this ordering by MAPE does not predict the relative quality of the corresponding SQD energies, suggesting that the proximity to $t_{2}^{\text{CCSD}}$ is not a reliable metric for LUCJ performance.

\subsection{Analysis of Total Energies}\label{subsection:energies}
\begin{figure}[H]
    \centering
    \includegraphics[width=0.75\linewidth]{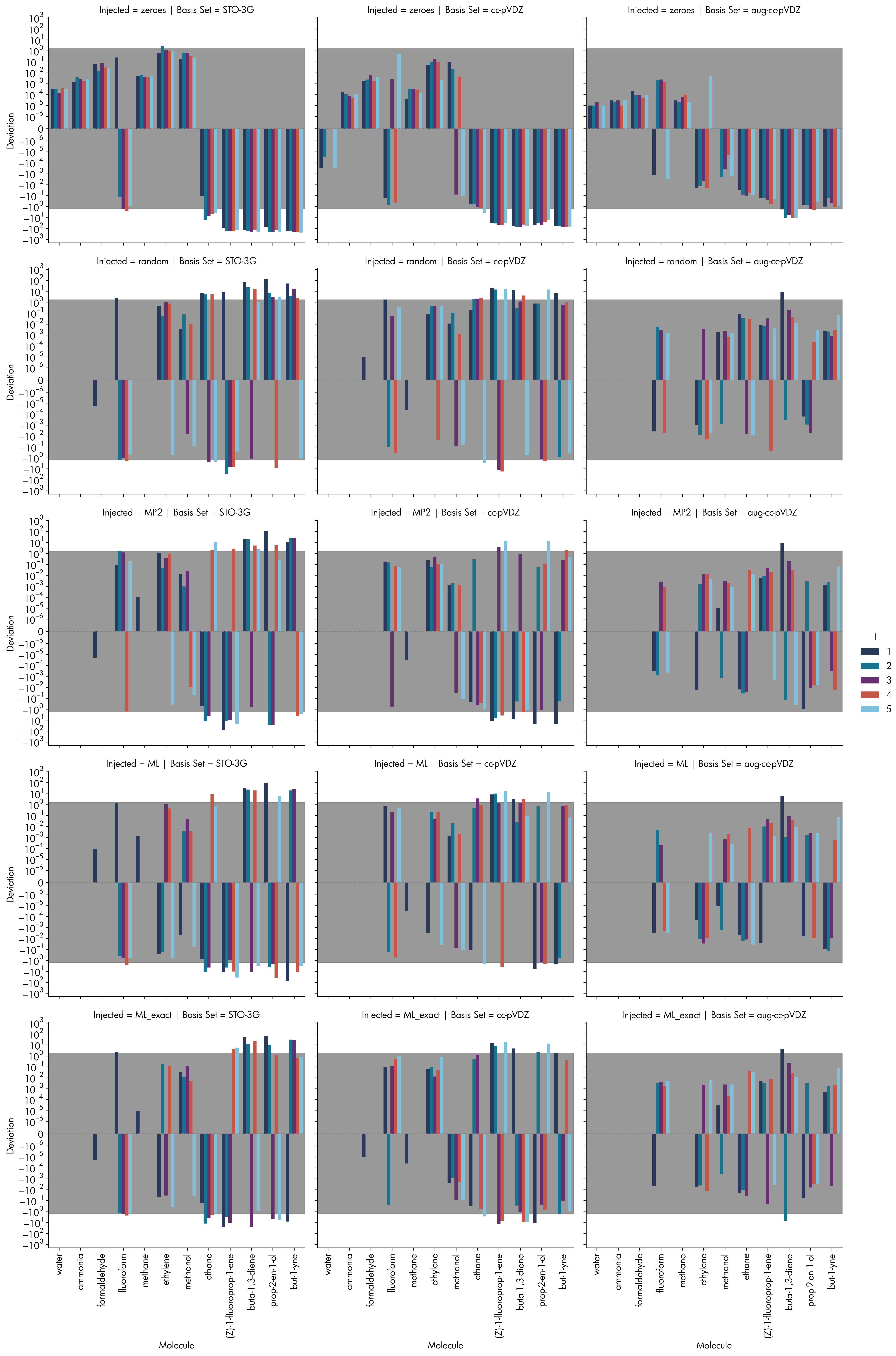}
    \caption{Deviations (in m$E_{\mathrm{h}}$) between the converged energies of LUCJ ans\"{a}tze initialized using various parameter injection strategies compared against the CCSD-initialized LUCJ ansatz, computed across a set of small organic molecules. Results are shown for five ansatz depths ($L = 1$--$5$, indicated by the hue) and three basis sets (STO-3G, cc-pVDZ, and aug-cc-pVDZ), arranged in a $5\times3$ grid by injection method (rows) and basis set (columns). The $y$-axis employs a symmetric logarithmic scale to accommodate the wide dynamic range of deviations observed across strategies and molecules. All deviations are reported relative to the CCSD-initialized ansatz energy, which by construction yields zero deviation (not shown).}
    \label{fig:DevEnergiesCCSDInitAll}
\end{figure}

Next, in Figure~\ref{fig:DevEnergiesCCSDInitAll}, we analyze the deviations in the total energies (in m$E_{\mathrm{h}}$) between the alternative initialization schemes and the standard CCSD initialization across the three basis sets (STO-3G, cc-pVDZ, and aug-cc-pVDZ) and for ansatz depths ranging from $L=1$ to $5$.
We note that if a bar is not present, the energy difference between the corresponding injection scheme and the CCSD initialization is zero (equivalent up to the sixth decimal place).
Across the different injection schemes, basis sets, and number of layers, three notable trends emerge.

First, increasing ansatz depth generally improves performance, with the exception of the zeroes initialization.
All alternative initializations except zeroes show an improvement as $L$ increases: the random initialization improves from 69.44\% ($L=1$) to 86.11\% ($L=5$) of values within chemical accuracy of CCSD, MP2 from 75\% to 83.33\%, ML from 75\% to 80.56\%, and ML\_exact from 72.22\% to 83.33\%.
In contrast, the zeroes initialization begins at 75\% for $L=1$ and decreases to 69.44\% at $L=5$.
The degradation of the zeroes initialization with increasing $L$ may reflect the fact that a zero-initialized ansatz amplifies the bias of identity-like operations.

Second, performance is strongly molecule-dependent. For all alternative initializations, regardless of basis set or ansatz depth, ammonia, methanol, water, methane, and formaldehyde are within chemical accuracy of the standard CCSD initialization. 
Of the remaining molecules, 98.89\% of ethylene, 86.67\% of fluoroform, 70\% of ethane, 64.44\% of but-1-yne, 55.56\% of prop-2-en-1-ol, 50\% of buta-1,3-diene, and 47.78\% of (Z)-1-fluoroprop-1-ene energies are within chemical accuracy of the CCSD reference. 
The poorer performance on larger molecules likely reflects the increased multireference character of these molecules, which places greater demands on both the expressivity of the LUCJ ansatz and the configurational diversity generated during SQD.

% Second, performance is strongly molecule-dependent.
% For all alternative initializations, regardless of basis set or ansatz depth, ammonia, methanol, water, methane, and formaldehyde are within chemical accuracy of the standard CCSD initialization.
% Of the remaining molecules, 98.89\% of ethylene, 86.67\% of fluoroform, 70\% of ethane, 64.44\% of but-1-yne, 55.56\% of prop-2-en-1-ol, 50\% of buta-1,3-diene, and 47.78\% of (Z)-1-fluoroprop-1-ene energies are within chemical accuracy of the CCSD reference.
% The poorer performance on larger molecules likely reflects the increased multireference character of these molecules, which places greater demands on both the expressibility of the LUCJ ansatz and the configurational diversity generated during SQD.

Third, agreement with the CCSD reference improves with basis set size.
The zeroes initialization improves from 55\% (STO-3G) to 88.33\% (aug-cc-pVDZ), random from 60\% to 98.33\%, MP2 from 60\% to 98.33\%, ML from 61.66\% to 98.33\%, and ML\_exact from 63.33\% to 96.66\%.
This trend is consistent with the expectation that larger basis sets provide increased accuracy, reducing sensitivity to the initial ansatz parameters.

These trends are important to note with respect to the MAPEs discussed in Section~\ref{subsection:amplitudes}.
Despite the $t_{2}^{\text{random}}$-amplitudes having the largest MAPEs relative to $t_{2}^{\text{CCSD}}$, the random initialization achieves competitive performance in terms of total energies within chemical accuracy, matching the best-performing MP2 and ML strategies at 98.33\% for the aug-cc-pVDZ basis set.
Similarly, the $t_{2}^{\text{zeroes}}$-amplitudes, which are closer to $t_{2}^{\text{CCSD}}$ in terms of MAPE than both $t_{2}^{\text{ML}}$ and $t_{2}^{\text{random}}$, yield the worst total energy agreement across all basis sets and values of $L$, confirming that amplitude-space proximity to $t_{2}^{\text{CCSD}}$ is not a reliable predictor of the quality of the resulting LUCJ energy.
This apparent disconnect indicates that the SQD optimization is relatively robust to the choice of initial $t_2$-amplitudes, provided the initialization does not impose a systematic bias on the ansatz parameters, which is precisely the case for the zeroes initialization.
Taken together, these results suggest that the primary driver of energy agreement is not the accuracy of the injected amplitudes themselves, but rather the ability of the resulting LUCJ parameters to place the ansatz in a favorable region of the variational landscape.
Notably, the $\Tilde{t}_{2}$-amplitudes, despite having the smallest MAPEs relative to $t_{2}^{\text{CCSD}}$, do not universally outperform $t_{2}^{\text{MP2}}$ or $t_{2}^{\text{ML}}$ in terms of energies within chemical accuracy, further confirming that amplitude proximity to the CCSD solution does not guarantee superior variational performance.

\subsection{Comparison with Classical Reference Methods}
Having established the robustness of SQD to initialization choice, we now assess the absolute accuracy of all six strategies relative to classical benchmarks.
Across all molecules, ansatz layers, and injection methods (including CCSD), the mean absolute errors (MAEs) between HCI and the LUCJ ans\"{a}tze are 70.76 m$E_{\mathrm{h}}$ (STO-3G), 20.10 m$E_{\mathrm{h}}$ (cc-pVDZ), and 1.01 m$E_{\mathrm{h}}$ (aug-cc-pVDZ).

This can be broken down further by injection method, where the STO-3G basis set has MAEs of 21.87 m$E_{\mathrm{h}}$ (zeroes), 83.81 m$E_{\mathrm{h}}$ (random), 79.53 m$E_{\mathrm{h}}$ (MP2), 79.16 m$E_{\mathrm{h}}$ (ML), 81.18 m$E_{\mathrm{h}}$ (ML\_exact), and 79.01 m$E_{\mathrm{h}}$ (CCSD).
For the cc-pVDZ basis set, we observe a similar trend: 6.13 m$E_{\mathrm{h}}$ (zeroes), 23.69 m$E_{\mathrm{h}}$ (random), 21.84 m$E_{\mathrm{h}}$ (MP2), 23.46 m$E_{\mathrm{h}}$ (ML), 22.83 m$E_{\mathrm{h}}$ (ML\_exact), and 22.68 m$E_{\mathrm{h}}$ (CCSD).
Regardless of injection choice, the aug-cc-pVDZ basis set has MAEs all within chemical accuracy of HCI, ranging from 0.27 m$E_{\mathrm{h}}$ (zeroes) to 1.24 m$E_{\mathrm{h}}$ (random).
We note that the MAEs reported here represent averages over molecules, ansatz layers, and SQD batches; batch-to-batch variance and shot noise contributions are not separately quantified. 
This omission represents a limitation of the present analysis.

We perform a similar analysis using the exact solution within the given active space provided by CASCI for all molecules with active spaces smaller than (20,20), i.e., excluding fluoroform, buta-1,3-diene, but-1-yne, prop-2-en-1-ol, and (Z)-1-fluoroprop-1-ene.
For the STO-3G, cc-pVDZ, and aug-cc-pVDZ basis sets, the MAEs between CASCI and the LUCJ methods are 1.33 m$E_{\mathrm{h}}$, 0.28 m$E_{\mathrm{h}}$, and 0.01 m$E_{\mathrm{h}}$, respectively. 
When broken down by injection method, the MAEs range from 0.85 m$E_{\mathrm{h}}$ (zeroes) to 1.91 m$E_{\mathrm{h}}$ (random) for the STO-3G basis set, 0.09 m$E_{\mathrm{h}}$ (zeroes) to 0.42 m$E_{\mathrm{h}}$ (random) for cc-pVDZ, and $2.17 \times 10^{-3}$ m$E_{\mathrm{h}}$ (zeroes) to $1.70 \times 10^{-2}$ m$E_{\mathrm{h}}$ (random) for aug-cc-pVDZ, with the remaining initializations falling between these bounds.
These results highlight that for small systems where exact diagonalization can be performed, the choice of initialization has a negligible effect on the recovered energy.
Furthermore, when compared against HCI for larger systems, the zeroes initialization offers the most consistent results across all basis sets, despite its poor performance relative to the CCSD reference observed in Section~\ref{subsection:energies}.
This finding underscores the importance of the choice of reference method when evaluating initialization strategies.
\section{Conclusion}
In this work, we have systematically analyzed the dependence of the SQD algorithm on the choice of LUCJ ansatz initialization across six strategies, twelve molecules, three basis sets, and five ansatz depths. 
Our results demonstrate that the SQD framework within the QCSC architecture is broadly robust to initialization choice: despite MAPEs between the alternative and CCSD $t_{2}$-amplitudes spanning up to ten orders of magnitude, most initializations recover total energies within chemical accuracy of the CCSD reference.
A central finding of this study is the apparent disconnect between the choice of initialization and the quality of the recovered electronic energy. 
The zeroes initialization, which has smaller MAPEs relative to CCSD than either the ML or random strategies, consistently yields the worst energy agreement across basis sets and ansatz depths. 
Conversely, random initialization, despite having MAPEs up to $10^{10}$\%, achieves competitive performance with the best-performing strategies at larger basis sets. 
This behavior is consistent with the interpretation that SQD's configuration recovery is the dominant driver of energy accuracy, and that initialization primarily determines the region of the variational landscape explored by the ansatz rather than directly setting the recovered energy.
We also demonstrate that DDCC-based initialization (ML and ML{\_}exact) provides a practical middle ground: it avoids the computational overhead of full CCSD while achieving energy agreement comparable to MP2 and random initializations. 
This suggests that DDCC-based initialization is a viable, scalable alternative for large-scale QCSC workflows where CCSD initialization becomes a computational bottleneck.
% Future work should examine the performance of these initialization strategies on larger systems where CASCI is intractable, and investigate the interplay between initialization choice and hardware noise levels. 
% Additionally, exploring whether DDCC models trained on small molecules transfer to larger active spaces remains an open and practically important question.

\section{Author Contributions}
GMJ: conceptualization, methodology, software, validation, formal analysis, investigation, data curation, writing - original draft, writing - review and editing, visualization. 
MJA: software, validation, formal analysis, investigation, data curation, writing - original draft, writing - review and editing. 
MOL: software, validation, formal analysis, investigation, data curation, writing - original draft, writing - review and editing.
HAJ: conceptualization, resources, writing - review and editing, supervision, project administration, funding acquisition.

\section{Conflicts of Interest}
The authors declare no competing financial interests.

\section{Acknowledgments}
We acknowledge the Government of Canada’s New Frontiers in Research Fund (NFRF), for grant NFRFE-2022-00226, and the Quantum Software Consortium (QSC), financed under grant \#ALLRP587590-23 from the National Sciences and Engineering Research Council of Canada (NSERC) Alliance Consortia Quantum Grants. 
We also acknowledge Mitacs Canada (grant IT46277) for the award and support for M.O.L. 
This research was enabled in part by computational support provided by IBM Quantum via the Quantum Software Consortium and PINQ2, along with access to classical resources through the Digital Research Alliance of Canada.

\section{Data and Software Availability}
All code and data used in this study are hosted on GitHub, free of charge at\\~\href{https://github.com/MSRG/LUCJSQDInitializationAnalysis}{https://github.com/MSRG/LUCJSQDInitializationAnalysis}.

\section{Supplementary Information}
Information regarding the SHAP feature analysis, machine learning details, and molecular structures can be found in the Supplementary Information.

\bibliography{achemso-demo}

\end{document}

\setcounter{page}{1}
\renewcommand{\thepage}{S-\arabic{page}}

\pagebreak
\tableofcontents
\newpage

\setcounter{figure}{0}
\renewcommand{\figurename}{Figure}
\renewcommand{\thefigure}{S\arabic{figure}}

\setcounter{table}{0}
\renewcommand{\tablename}{Table}
\renewcommand{\thetable}{S\arabic{table}}

\setcounter{section}{0}
\renewcommand{\thesection}{S\arabic{section}}

% \pagebreak
% \section{Active space}\label{section:activespace}

% \begin{table}[h!]
%     \centering
%     \begin{tabular}{ccc}
%         \hline
%         Molecule & n\textunderscore orbitals & n\textunderscore electrons  \\
%         \hline
%         water          & 7  & 10         \\
%         ammonia        & 8  & 10        \\ 
%         methane        & 9  & 10      \\
%         formaldehyde   & 12 & 16           \\
%         ethylene       & 14 & 16        \\
%         ethane         & 16 & 18           \\
%         methanol       & 14 & 18        \\
%         GDB04\textunderscore 53      & 26 & 30        \\
%         GDB04\textunderscore 49      & 26 & 30        \\
%         GDB04\textunderscore 33      & 26 & 32      \\ 
%         GDB04\textunderscore 65      & 25 & 32        \\
%         GDB04\textunderscore 5       & 21 & 34      \\
%         \hline
%     \end{tabular}
%     \caption{Active spaces for the different molecules in the datasets.}
%     \label{tab:benchmark}
% \end{table}

\pagebreak
\section{SHAP Feature analysis}\label{section:feature_analysis}

\begin{table}[h]
  \centering
  \renewcommand{\arraystretch}{1.5} 
  \begin{tabular}{|l|l|l|}
    \hline
    \textbf{Variable} & \textbf{Symbol} & \textbf{Description} \\
    \hline
    doublecheck & $\Braket{ij || ab}$ & two-electron integral\\
    orbdiff&$(\epsilon_i + \epsilon_j - \epsilon_a - \epsilon_b)$ & MP2 denominator\\
    diag&$\delta_{ab}$ & whether the two virtual orbitals are identical \\
    t2start&${t_{ij}^{ab}}_{\text{MP2}}$ & initial MP2 amplitude\\
    t2mag&$\log (| {t_{ij}^{ab}}_{\text{MP2}}|)$  & magnitude of initial MP2 amplitude\\ [1ex] 
    \hline
  \end{tabular}
  \caption{Description of the top five features}
  \label{tab:SHAP_table}
\end{table}

\begin{figure}[htbp]
    \centering
    \includegraphics[width=\linewidth]{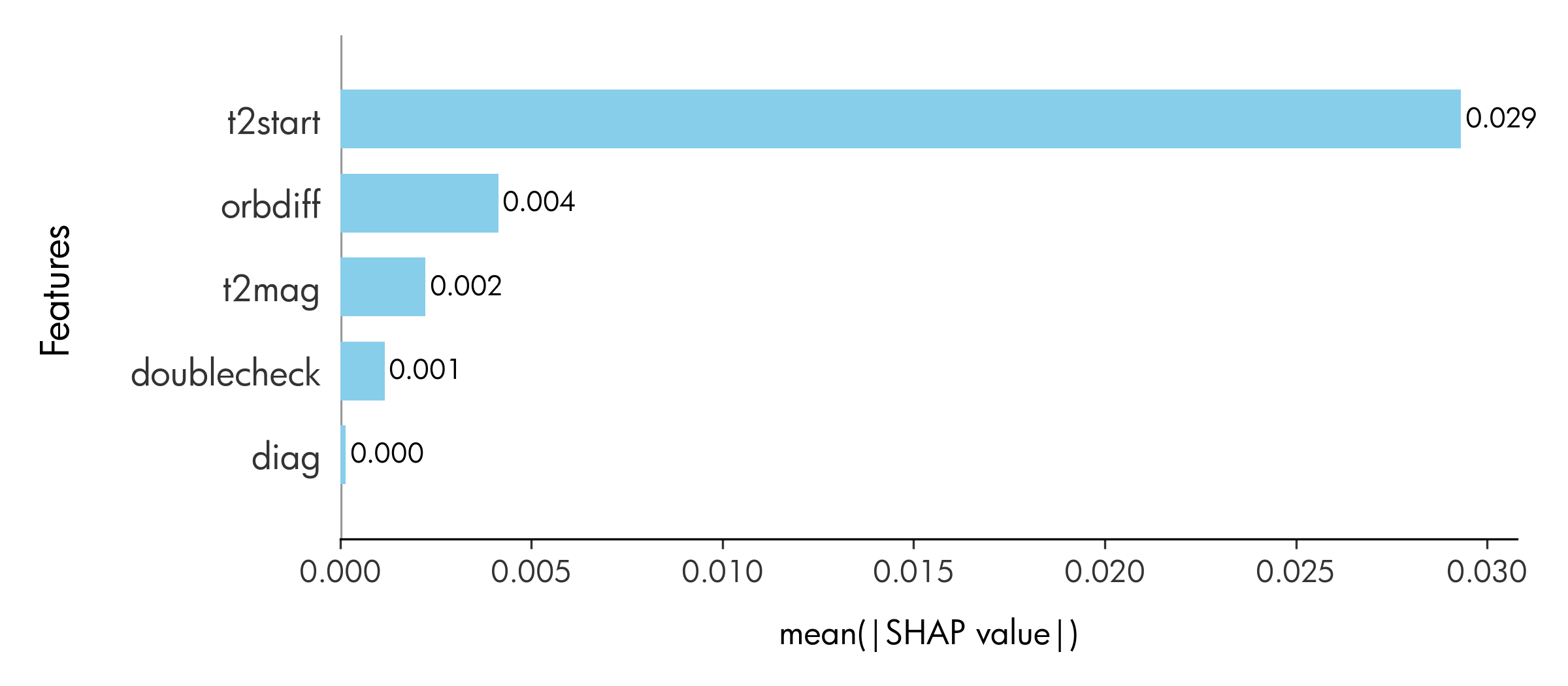}
    \caption{SHAP analysis of the top five most important features to guide the prediction of $t_2$. The greater the magnitude of the mean SHAP value, the more important the feature is.}
    \label{fig:shap-analysis}
\end{figure}

\begin{figure}[htbp]
    \centering
    \includegraphics[width=0.8\linewidth]{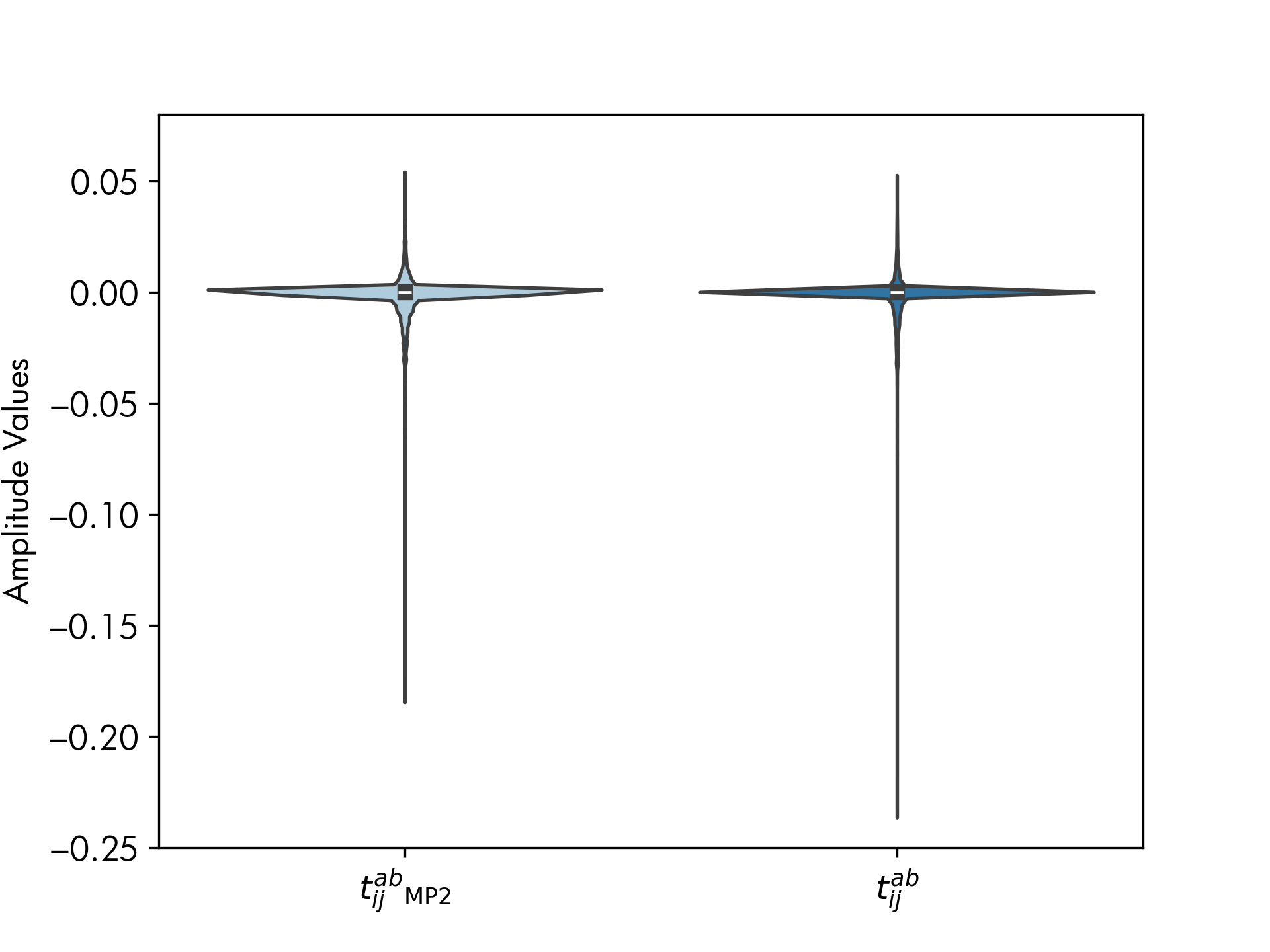}
    \caption{Violin plot of the values of t2start (left) and t2 (right).}
    \label{fig:violin}
\end{figure}

\begin{figure}[htbp]
    \centering
    \includegraphics[width=\linewidth]{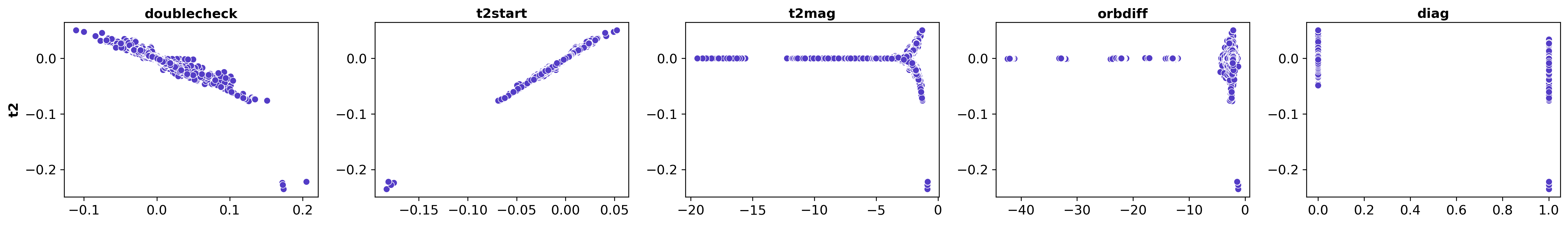}
    \caption{Relationship between the target variable, $t_2$, and the top five features. Points that lie on a perfect diagonal indicate a strong correlation for this feature.}
    \label{fig:corr_per_feature}
\end{figure}

\begin{figure}[htbp]
    \centering
    \includegraphics[width=\linewidth]{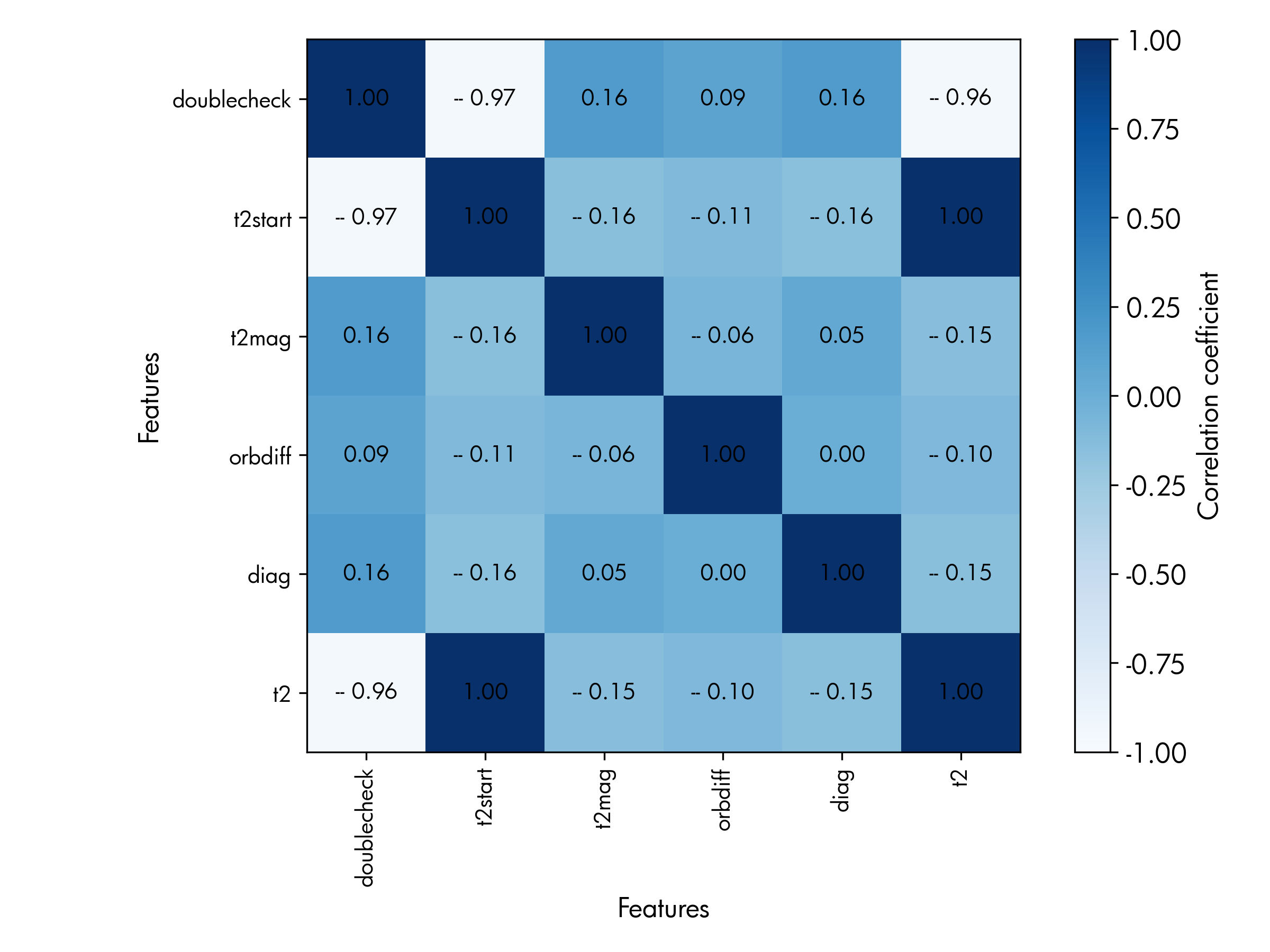}
    \caption{Correlation matrix between the top five features and $t_2$. Darker blue indicates a strong positive correlation, while lighter shading (approaching white) indicates a strong negative correlation. This matrix is symmetric, meaning the upper triangle is the mirror image of the lower triangle.}
    \label{fig:corr_top5}
\end{figure}

% \begin{figure}[htbp]
%     \centering
%     \includegraphics[width=\linewidth]{figures/suppinfo/tsne.png}
%     \caption{Embedding representation of the molecular feature inputs based on the top five features. Each point represents a single molecule's feature vector projected into the two-dimensional t-SNE space. Molecules with similar feature vectors tend to cluster together. The organized pattern of the embeddings indicate a non-linear relationship between features. The overlap in space between the training and testing data suggests that the feature distributions between the two sets are comparable and representative of the true distribution.}
%     \label{fig:tsne}
% \end{figure}

\pagebreak
\section{Machine Learning}\label{section:ml}

\begin{table}[H]
  \centering
  \renewcommand{\arraystretch}{1.5} 
  \begin{tabular}{|c|c|c|c|c|c|}
    \hline
    \textbf{Method} & \textbf{\textit{N}} & \textbf{reg\textunderscore lambda} & \textbf{reg\textunderscore alpha} & \textbf{max\textunderscore depth} & \textbf{n\textunderscore estimators} \\
    \hline
    \multirow{6}{*}{\textbf{STO-3G}} 
      & 10  & $10^{-1}$  & $ 10^{-3}$  & 15 & 200 \\ \cline{2-6}
      & 20  & $10^{-3}$  & $ 10^{-6}$  & 10 & 400 \\ \cline{2-6}
      & 40  & $ 10^{-6}$  & $ 10^{-6}$  & 10 & 500 \\ \cline{2-6}
      & 60  & $ 10^{-6}$  & $ 10^{-6}$  & 10 & 500 \\ \cline{2-6}
      & 80  & $ 10^{-6}$  & $ 10^{-6}$  & 15 & 500 \\ \cline{2-6}
      & 100 & $ 10^{-6}$  & $ 10^{-6}$  & 15 & 400 \\ \hline
    \hline
    \multirow{6}{*}{\textbf{cc-pVDZ}} 
      & 10  & $ 10^{-3}$  & $ 10^{-6}$  & 20 & 400 \\ \cline{2-6}
      & 20  & $ 10^{-3}$  & $ 10^{-6}$  & 20 & 400 \\ \cline{2-6}
      & 40  & $ 10^{-6}$  & $ 10^{-6}$  & 20 & 500 \\ \cline{2-6}
      & 60  & $ 10^{-3}$  & $ 10^{-6}$  & 20 & 500 \\ \cline{2-6}
      & 80  & $ 10^{-6}$  & $ 10^{-6}$  & 20 & 400 \\ \cline{2-6}
      & 100 & $ 10^{-6}$  & $ 10^{-6}$  & 20 & 400 \\ \hline
    \hline
    \multirow{6}{*}{\textbf{aug-cc-pVDZ}} 
      & 10  & $ 10^{-3}$  & $ 10^{-6}$  & 20 & 500 \\ \cline{2-6}
      & 20  & $ 10^{-3}$  & $ 10^{-6}$  & 20 & 500 \\ \cline{2-6}
      & 40  & $ 10^{-6}$  & $ 10^{-6}$  & 20 & 500 \\ \cline{2-6}
      & 60  & $ 10^{-6}$  & $ 10^{-6}$  & 20 & 500 \\ \cline{2-6}
      & 80  & $ 10^{-3}$  & $ 10^{-6}$  & 20 & 500 \\ \cline{2-6}
      & 100 & $ 10^{-3}$  & $ 10^{-6}$  & 20 & 500 \\
    \hline
  \end{tabular}
  \caption{Optimized hyperparameters obtained via grid search for the XGBoost models, shown as the number of training points (\textbf{\textit{N}}) is progressively increased. \textbf{reg{\_}lambda} and \textbf{reg{\_}alpha} denote the L2 and L1 regularization parameters, respectively.}
  \label{tab:hyperparameters}
\end{table}

\pagebreak
\section{Coordinates for the GDB-11 molecules used in this study.}\label{section:cords1}

\begin{verbatim}
GDB04_05
F      1.261484     -0.405551      0.050194 
C     -0.000242      0.070537     -0.013167 
F     -0.650370     -0.605250     -0.984390 
F     -0.606936     -0.204607      1.161076 
H     -0.003936      1.144871     -0.213713

GDB04_33
O     -1.928118     -0.114744      0.178876 
C     -0.671407     -0.765721      0.264554 
C      0.284879      0.150282      0.953172 
C      1.413808      0.622382      0.408653 
H     -1.767606      0.745694     -0.244961 
H     -0.344695     -1.031447     -0.745713 
H     -0.795495     -1.687137      0.840773 
H      0.027245      0.434591      1.971228 
H      1.715326      0.362710     -0.601521 
H      2.066064      1.283390      0.971827 

GDB04_49
C     -1.058917     -0.590121     -0.502810 
C      0.095088      0.393251     -0.599948 
C      0.899775      0.394851      0.629317 
C      1.559705      0.398731      1.632574 
H     -0.696202     -1.611412     -0.343086 
H     -1.725428     -0.337209      0.328999 
H     -1.647241     -0.578510     -1.425782 
H      0.726442      0.129767     -1.455754 
H     -0.298457      1.398615     -0.786497 
H      2.145235      0.402037      2.522987 

GDB04_53
C     -1.043781      0.375335      1.269531 
C     -0.416543     -0.312814      0.308761 
C      0.416541      0.312814     -0.688124 
C      1.043779     -0.375334     -1.648895 
H     -1.661571     -0.137665      1.999907 
H     -0.961833      1.454076      1.355928 
H     -0.534094     -1.393801      0.269412 
H      0.534091      1.393801     -0.648773 
H      1.661582      0.137664     -2.379262 
H      0.961830     -1.454075     -1.735295 

GDB04_65
C     -1.007357     -0.507621     -0.137822 
C      0.112792      0.215978      0.531784 
C      1.251976      0.555729     -0.072580 
F      1.491289      0.268016     -1.364635 
H     -0.791187     -0.731217     -1.187146 
H     -1.917773      0.098316     -0.099357 
H     -1.201645     -1.452609      0.378651 
H     -0.024138      0.475112      1.577742 
H      2.086041      1.078295      0.373363 
\end{verbatim}

\pagebreak
\section{Coordinates for the benchmark conformer of each small molecule}\label{section:cords2}
\begin{verbatim}
ammonia157
H      1.466280     0.999842    -0.865760
H      0.416028     0.110556     0.002083
H      1.751150     0.708500     0.691277
N      1.010810     0.931390     0.037977

methane50
H     -0.014612     0.692188    -0.266503
H      1.724740     0.343450    -0.481446
H      1.165000     2.026840    -0.333225
H      1.121960     0.941367     1.081300
C      0.999002     0.999952     0.000058

ethylene42
C     -1.141370     1.730050     0.795470
C     -0.336198     0.999753     0.000194
H     -1.303320     1.450720     1.826430
H     -1.638850     2.615320     0.417248
H      0.164333     0.113727     0.369845
H     -0.166856     1.273140    -1.033050

ethane28
H      0.247375    -0.954156    -0.673398
C      0.998505     0.999338    -0.001117
H      2.054340     0.726303    -0.024416
H      0.772656     1.531530    -0.926205
H      0.326639    -0.789683     1.079760
C      0.103445    -0.255274     0.154709
H     -0.952837     0.011962     0.178581
H      0.852722     1.692400     0.828762

water183
O     10.000000    10.000000    10.000000
H     10.554100    10.192500    10.759600
H     10.528500    10.033700     9.201820

formaldehyde138
O      1.013160     0.999902    -0.000000
H      2.566200     0.336199     1.146540
H      2.818880     1.911760     0.237052
C      2.149080     1.082500     0.469005

methanol22
H      0.874204     1.809430    -0.690054
H      1.897770     0.406931    -0.335946
H      0.152751     0.335913    -0.028639
H      1.380530     0.748667     1.914430
O      1.245810     1.474360     1.321060
C      1.032440     0.980422     0.004686

\end{verbatim}

\pagebreak
\section{Coordinates for training-set molecules}\label{section:cords3}
\begin{verbatim}
ethane179
C     -1.569340     0.508349     0.176131
C     -0.336198     0.999753     0.000194
H     -2.385630     1.134650     0.505486
H     -1.777720    -0.538357     0.010019
H     -0.125273     2.044220     0.164338
H      0.476973     0.371545    -0.318568

methane95
H      0.968367     0.023777    -0.487154
H      1.049940     1.789540    -0.754862
H      1.877100     1.045500     0.647698
H      0.090896     1.128690     0.590428
C      0.999002     0.999952     0.000058

water113
O     10.000000    10.000000    10.000000
H     10.556300     9.781410    10.748700
H      9.895100    10.950800     9.947690

ethylene66
C      0.200112     0.677424    -1.184060
C     -0.336198     0.999753     0.000194
H     -0.096978    -0.225864    -1.696310
H      0.947540     1.300680    -1.659000
H     -1.081850     0.379945     0.476765
H     -0.033392     1.898720     0.512546

ammonia72
N      1.014200     0.981266    -0.008802
H      0.775528     0.695689    -0.948177
H      1.286960     0.137721     0.482327
H      0.146052     1.292160     0.411335

methane175
C      0.998449     0.859639     0.016817
H     -0.091029     1.000000     0.000028
H      1.451370     1.561760    -0.683193
H      1.243590    -0.162351    -0.276475
H      1.369580     1.058030     1.021440

methane23
H      1.738700     0.590117    -0.691851
H      0.679003     1.978230    -0.360183
H      0.133736     0.332754     0.055492
H      1.450130     1.091550     0.988552
C      0.999002     0.999952     0.000058

ethane70
H     -0.865750    -0.126503    -0.301462
H      0.585123    -1.041740    -0.689879
C      0.998505     0.999338    -0.001117
H      1.018780     1.450240    -0.994071
H      0.145275    -0.761976     0.991261
C      0.167445    -0.307275     0.000224
H      2.030150     0.811915     0.299840
H      0.577810     1.731630     0.689114

methane7
H      0.632693     0.207603    -0.655809
H      0.830962     1.965270    -0.478397
H      2.069250     0.867538     0.188568
C      0.999002     0.999952     0.000058
H      0.469899     0.958335     0.955467

ethane116
C      0.782787     1.290140     0.686643
C     -0.336198     0.999753     0.000194
H      1.708020     0.770220     0.483761
H      0.791450     2.049730     1.454230
H     -0.341408     0.242636    -0.768599
H     -1.262950     1.521660     0.194466

ammonia16
H      0.824326     0.731782    -0.954161
H      2.001520     1.146690     0.077067
H      0.808413     0.163800     0.559246
N      1.005010     0.988975     0.006513

ethylene41
C     -0.175586    -0.308949    -0.241463
C     -0.336198     0.999753     0.000194
H     -0.117766    -0.693422    -1.250300
H     -0.111633    -1.026930     0.565688
H     -0.415750     1.714950    -0.805755
H     -0.401857     1.373350     1.011900

ammonia196
H      0.066630     1.232420    -0.152542
H      1.428660     0.770890    -0.896815
H      1.507010     1.727920     0.401747
N      1.018430     0.931384     0.012858

methanol44
H      1.337150    -0.036259    -0.253737
H      0.560342     1.490070    -1.818190
O      1.117180     1.827020    -1.131650
H      0.012766     0.950690     0.393572
H      1.690820     1.346630     0.795083
C      1.032440     0.980422     0.004686

ethylene17
C     -1.607380     0.637613    -0.234457
C     -0.336198     0.999753     0.000194
H     -2.242260     1.297750    -0.805789
H     -2.000610    -0.302896     0.110034
H      0.042468     1.948610    -0.363753
H      0.347921     0.361084     0.551339

methanol49
H     -0.297283     1.218720    -1.401420
H      1.737180     1.723780    -0.375247
O      0.210292     0.500708    -1.050570
H      1.604980     0.157492     0.437597
H      0.431121     1.436750     0.793863
C      1.032440     0.980422     0.004686

water179
O     10.000000    10.000000    10.000000
H      9.457500     9.438410     9.442980
H      9.517080    10.216800    10.798300

formaldehyde196
H     -0.473694     0.718197    -1.366310
C     -0.113903     1.197670    -0.455124
O      1.013160     0.999902    -0.000000
H     -0.832384     1.865660     0.021242

ethylene52
C     -0.959766     0.221905    -0.902804
C     -0.336198     0.999753     0.000194
H     -0.593169     0.115886    -1.919740
H     -1.874650    -0.299987    -0.643929
H      0.574541     1.533620    -0.254151
H     -0.721649     1.100690     1.007950

ammonia108
H      0.125093     0.981594    -0.487076
H      1.724480     1.168660    -0.634487
H      0.956142     1.769180     0.659216
N      0.990577     0.972225     0.034869

formaldehyde52
H      2.423400     2.452820    -0.224602
H      2.999790     0.715317    -0.373769
O      1.013160     0.999902    -0.000000
C      2.162550     1.394080    -0.202808

methanol64
H      0.012807     0.615110    -0.135019
H      1.212430     1.781430    -0.714772
H      1.727030     0.165524    -0.206986
H      1.026480     0.771657     1.944470
O      1.219630     1.462920     1.325520
C      1.032440     0.980422     0.004686

ethane109
C     -0.794295    -0.255864     0.090340
C     -0.336198     0.999753     0.000194
H     -0.136291    -1.052880     0.389633
H     -1.821800    -0.505623    -0.120027
H      0.694374     1.250560     0.228068
H     -1.003760     1.802920    -0.289196

methanol41
O     -0.021948     0.031952    -0.066315
H      1.413940     1.195450    -0.995435
H     -0.355149    -0.132933     0.804198
H      1.854670     0.592473     0.608410
H      0.684462     1.915800     0.447491
C      1.032440     0.980422     0.004686

water42
O     10.000000    10.000000    10.000000
H      9.652970    10.787300     9.571070
H      9.277390     9.458630    10.324000

ethane187
C      0.196661    -0.138069    -0.484209
C     -0.336198     0.999753     0.000194
H     -0.068487    -1.096990    -0.046284
H      0.898384    -0.144724    -1.307130
H     -1.024970     0.976996     0.832961
H     -0.071763     1.962550    -0.413736

methane76
H      1.753430     1.666300    -0.420508
H      1.042550     0.037787    -0.508390
H      0.009160     1.439420    -0.135094
H      1.193930     0.859980     1.062790
C      0.999002     0.999952     0.000058

ammonia156
H      0.716090    -0.005994    -0.198371
H      0.571927     1.550110    -0.632161
H      0.591746     1.138040     0.933026
N      1.014670     0.932638     0.035670

methanol37
H      1.409370    -0.022366    -0.205810
H      0.520607     1.356880    -1.840500
O      1.055590     1.770560    -1.176810
H      1.655820     1.427530     0.782192
H      0.015183     0.895439     0.393323
C      1.032440     0.980422     0.004686

ethane155
C     -1.479020     0.620488    -0.605560
C     -0.336198     0.999753     0.000194
H     -2.287180     0.192748    -0.035172
H     -1.621370     0.726551    -1.671330
H     -0.197584     0.891409     1.069570
H      0.472568     1.429010    -0.574042

methane93
H      1.547300     0.198151    -0.491649
H      0.257553     1.414700    -0.688839
H      0.495723     0.592907     0.881003
H      1.702570     1.781000     0.295859
C      0.999002     0.999952     0.000058

ethane172
C     -0.441079     1.760530    -1.095580
C     -0.336198     0.999753     0.000194
H     -1.268100     2.459820    -1.240710
H      0.332380     1.741450    -1.867890
H     -1.073500     1.012410     0.759648
H      0.488288     0.341899     0.130925

formaldehyde170
O      1.013160     0.999902    -0.000000
H      1.584100     0.003450     1.688290
C      0.881665     0.665265     1.179530
H      0.051713     1.009980     1.798500

formaldehyde101
H     -0.519184     0.591780    -1.305490
H     -0.301274     2.323370    -0.824518
O      1.013160     0.999902    -0.000000
C      0.085512     1.310100    -0.750858

ethylene45
C     -1.280400     1.638080    -0.690589
C     -0.336198     0.999753     0.000194
H     -2.333470     1.639250    -0.448955
H     -0.955165     2.172890    -1.574520
H     -0.500573     0.385338     0.875405
H      0.685178     1.020260    -0.356502

ammonia174
H      0.705321     0.646383    -0.901150
H      0.359574     0.515093     0.672279
H      0.905320     1.919130     0.074017
N      1.020900     0.915991     0.020654

formaldehyde88
O      1.013160     0.999902    -0.000000
H     -0.923183     0.931365     0.633519
H      0.332265     0.151025     1.725550
C      0.127803     0.687924     0.798257

methane193
C     -1.166190     0.867083     0.145774
H     -0.091029     1.000000     0.000028
H     -1.373470    -0.188962     0.315118
H     -1.707350     1.207170    -0.733398
H     -1.487380     1.440560     1.012550

ammonia160
H      1.445130     1.076160    -0.860271
H      0.017969     0.908114    -0.114733
H      1.203080     1.758880     0.580982
N      1.016830     0.924428     0.042846

ethylene47
C      0.020605    -0.270803    -0.237990
C     -0.336198     0.999753     0.000194
H      0.845155    -0.532746    -0.902161
H     -0.537206    -1.075430     0.238999
H      0.190637     1.840830    -0.431833
H     -1.176580     1.213660     0.651605

water30
O     10.000000    10.000000    10.000000
H      9.339420     9.325240     9.841570
H     10.511400     9.772010    10.777800

ammonia50
N      1.003220     0.991632    -0.007932
H      0.996759     1.899840    -0.454275
H      1.027690     0.309088    -0.752192
H      1.890000     0.921948     0.472557

water131
O     10.000000    10.000000    10.000000
H      9.288330     9.404240    10.238900
H      9.768070    10.896500    10.251500

water180
O     10.000000    10.000000    10.000000
H     10.353800     9.740690     9.148850
H     10.630100     9.781090    10.687800

methane101
C     -0.961005     0.670627     0.610523
H     -0.091029     1.000000     0.000028
H     -0.666935     0.503482     1.655600
H     -1.363130    -0.260786     0.213992
H     -1.770710     1.398330     0.668664

formaldehyde2
H      0.929962     2.627410    -1.197580
C      1.188050     1.580000    -1.080110
O      1.013160     0.999902    -0.000000
H      1.775700     1.152680    -1.898980

water93
O     10.000000    10.000000    10.000000
H      9.652970    10.787300     9.571070
H      9.277390     9.458630    10.324000

formaldehyde126
O      1.013160     0.999902    -0.000000
H      1.932180     2.600430    -0.861840
H      2.496240     0.937705    -1.399700
C      1.826300     1.519290    -0.765218

ammonia52
N      1.004460     0.993965    -0.009081
H      1.034870     1.776330    -0.651941
H      1.058960     0.152228    -0.568217
H      1.866330     1.033200     0.518759

ethane101
C     -1.597100     0.566335    -0.149810
C     -0.336198     0.999753     0.000194
H     -1.994050     0.148724    -1.085040
H     -2.258720     0.543832     0.729738
H      0.329392     0.984804    -0.854184
H      0.019960     1.389740     0.931070

ethylene33
C     -1.579210     0.504679     0.125100
C     -0.336198     0.999753     0.000194
H     -2.239570     0.415911    -0.720618
H     -1.928630     0.180168     1.085730
H      0.031686     1.331970    -0.958933
H      0.323469     1.084410     0.855227

ethane20
C      0.998505     0.999338    -0.001117
H      0.894860     0.155154    -0.684450
H      0.634089     1.890910    -0.513186
H      0.581017    -0.134999     1.830740
H     -0.843624     0.614751     1.121380
C      0.218415     0.756316     1.312830
H      0.320844     1.601120     1.999810
H      2.062400     1.140910     0.194906

water124
O     10.000000    10.000000    10.000000
H     10.434300     9.763800     9.179110
H     10.395700     9.512660    10.727400

methanol3
H      0.950719     0.535044    -1.895950
O      0.418163     1.011610    -1.284340
H      2.066890     1.378860    -0.041967
H      1.070190    -0.030572     0.403008
H      0.464397     1.605450     0.698198
C      1.032440     0.980422     0.004686

ammonia172
H      0.061714     1.213270    -0.101549
H      1.409830     0.800846    -0.905741
H      1.507160     1.699240     0.430101
N      1.020360     0.913145     0.020620

ethylene88
C     -1.404780     0.200799     0.129374
C     -0.336198     0.999753     0.000194
H     -1.354810    -0.689459     0.738889
H     -2.338040     0.422027    -0.368850
H      0.593982     0.771692     0.499709
H     -0.366964     1.896380    -0.604316

ethylene35
C      0.908092     0.507843    -0.109736
C     -0.336198     0.999753     0.000194
H      1.091550    -0.553677    -0.005928
H      1.750280     1.159600    -0.291513
H     -1.181680     0.354912     0.191643
H     -0.527305     2.060290    -0.094851

methanol100
H      0.319294     1.417470    -0.697653
H      1.686180     0.299824    -0.544255
H     -0.184596     0.891470     1.523980
H      1.646250     1.784510     0.418442
O      0.359095     0.283205     1.043180
C      1.032440     0.980422     0.004686

ammonia63
N      1.008030     0.988269    -0.011085
H      0.852502     0.554434    -0.909578
H      0.165303     1.500950     0.206128
H      1.063320     0.230335     0.656898

methane64
H      0.812348     0.110668    -0.602853
H      1.530710     1.746780    -0.596158
H      1.606190     0.728447     0.867352
H      0.047454     1.418570     0.327049
C      0.999002     0.999952     0.000058

water24
O     10.000000    10.000000    10.000000
H      9.284360    10.388100     9.497330
H      9.653960     9.310900    10.568500

water103
O     10.000000    10.000000    10.000000
H     10.360400     9.285150     9.465770
H     10.737700    10.116700    10.609700

ethane60
H      2.311020     1.215410    -1.736440
C      2.063540     1.753830    -0.820615
H      2.985320     1.878530    -0.251077
C      0.998505     0.999338    -0.001117
H      1.713720     2.747330    -1.105030
H      0.078077     0.875537    -0.572871
H      0.752347     1.538920     0.914491
H      1.352040     0.006650     0.280988

ammonia56
N      1.010980     0.994173    -0.011893
H      0.787235     0.823346    -0.983025
H      1.387080     1.931270     0.036199
H      1.775940     0.375513     0.224360

ammonia188
H      0.775005     0.682441    -0.933105
H      1.773160     1.594980    -0.042692
H      0.223535     1.429220     0.392534
N      1.016080     0.927080     0.017002

methanol67
H      0.443102     1.721010    -0.539857
H      1.473820     0.296583    -0.722649
H      0.724139    -0.369530     1.378830
H      1.832490     1.504220     0.532024
O      0.207239     0.278220     0.921016
C      1.032440     0.980422     0.004686

ethane71
H     -0.782140    -0.218000    -0.427835
H      0.640858    -1.165640    -0.021621
C      0.998505     0.999338    -0.001117
H      1.339150     1.068360    -1.035400
H     -0.242848    -0.318911     1.245240
C      0.101248    -0.246674     0.211817
H      1.880730     0.968142     0.640044
H      0.452315     1.914320     0.233375

ammonia13
H      1.083350     1.598530    -0.794240
H      1.041500     0.036845    -0.347844
H      1.840740     1.114670     0.546884
N      0.997282     0.983706     0.006350

ethane3
H     -0.036180     1.273010    -0.219549
C      0.998505     0.999338    -0.001117
H      1.615430     0.859237    -2.094560
H      2.962790     1.131640    -0.947134
C      1.924270     1.366600    -1.174300
H      1.909760     2.438480    -1.398490
H      1.041410    -0.067687     0.192944
H      1.272940     1.529410     0.914309

methanol35
H      0.377653     1.178050    -0.846436
H      1.865770     0.365930    -0.340470
H      0.914042     0.127028     1.753130
O      0.323058     0.319612     1.039090
H      1.427980     1.933130     0.361863
C      1.032440     0.980422     0.004686

ammonia29
N      0.995545     0.987045    -0.000157
H      0.443836     1.186390    -0.822490
H      1.859410     0.574380    -0.326114
H      1.239100     1.884280     0.396658

ethane167
C      0.911443     1.490270    -0.043382
C     -0.336198     0.999753     0.000194
H      1.095580     2.552160     0.037999
H      1.766940     0.835588    -0.153650
H     -1.191490     1.648110     0.113946
H     -0.519707    -0.063143    -0.077552

ammonia195
H      0.591485     0.851345    -0.899838
H      0.269262     1.152770     0.655265
H      1.342940     0.004941     0.258816
N      1.019690     0.932496     0.012417

ethylene69
C      0.953350     1.089120     0.364967
C     -0.336198     0.999753     0.000194
H      1.366720     2.015150     0.739799
H      1.614140     0.242059     0.278325
H     -1.001170     1.852960     0.074084
H     -0.754188     0.077840    -0.381393

formaldehyde15
O      1.013160     0.999902    -0.000000
H      1.682260     2.183450    -1.517490
H      2.952690     1.477980    -0.394507
C      1.895030     1.562130    -0.646936

formaldehyde133
O      1.013160     0.999902    -0.000000
H      2.682370     0.922862    -1.168750
C      2.123840     1.391700    -0.357810
H      2.627620     2.237570     0.111333

formaldehyde22
O      1.013160     0.999902    -0.000000
H      0.182646     0.785363     1.846980
H      1.918090     0.218450     1.649710
C      1.038310     0.662623     1.182740

ethylene81
C      0.887720     1.536700    -0.149272
C     -0.336198     0.999753     0.000194
H      1.058180     2.600690     0.008889
H      1.740560     0.951257    -0.460531
H     -1.183340     1.606730     0.292622
H     -0.511289    -0.055624    -0.162712

methane40
H      0.824848     1.890130    -0.607588
H      1.955680     0.557820    -0.276946
H      0.194547     0.280585    -0.170657
H      1.018570     1.271980     1.054770
C      0.999002     0.999952     0.000058

formaldehyde135
O      1.013160     0.999902    -0.000000
H      2.531400     0.148884     1.066590
C      2.101900     1.023830     0.576274
H      2.713670     1.924730     0.636975

formaldehyde32
C     -0.162769     1.128440    -0.339271
H     -0.473029     1.184140    -1.383280
O      1.013160     0.999902    -0.000000
H     -0.976381     1.195910     0.383349

formaldehyde94
H     -0.544066     1.191420    -1.302850
C     -0.181880     1.097980    -0.278324
O      1.013160     0.999902    -0.000000
H     -0.963450     1.097750     0.482480

ethane184
C     -1.308740     0.084737    -0.097942
C     -0.336198     0.999753     0.000194
H     -1.080900    -0.985860    -0.107328
H     -2.346890     0.384601    -0.169587
H      0.698974     0.702939     0.090401
H     -0.562720     2.052540     0.008608

ammonia92
N      1.008060     0.985274    -0.012561
H      0.607115     0.759255    -0.912929
H      1.929590     1.358960    -0.197838
H      0.456574     1.748040     0.357009

ethylene49
C      0.546169     0.411050     0.829684
C     -0.336198     0.999753     0.000194
H      1.369880    -0.175567     0.453269
H      0.460877     0.524931     1.901870
H     -0.254976     0.893102    -1.071550
H     -1.145060     1.606810     0.393493

ethylene16
C     -0.859356     0.313873    -1.028470
C     -0.336198     0.999753     0.000194
H     -1.904680     0.394727    -1.282340
H     -0.254115    -0.364613    -1.616780
H     -0.940935     1.672390     0.597187
H      0.711874     0.900282     0.254687

methanol39
H      0.880959     1.370420    -1.004240
H      1.727630     0.140938    -0.054241
H      1.702060     1.649550     1.711310
O      1.555340     1.999930     0.843970
H      0.073949     0.616273     0.378927
C      1.032440     0.980422     0.004686

methanol10
H      1.667750     0.206775    -0.431909
H      0.156954     1.104070    -0.633182
H      0.184986    -0.194333     1.313140
H      1.590580     1.925350     0.004075
O      0.671377     0.617391     1.329430
C      1.032440     0.980422     0.004686

methanol68
H      0.133782     1.991070    -1.399490
O      0.981356     1.585450    -1.280430
H      1.981120     0.454892     0.130003
H      0.222040     0.257675     0.126161
H      0.951886     1.730610     0.794912
C      1.032440     0.980422     0.004686

water59
O     10.000000    10.000000    10.000000
H     10.734000     9.741810    10.557000
H     10.333800    10.354300     9.175580

formaldehyde53
H      2.100220     2.251190    -1.182710
O      1.013160     0.999902    -0.000000
H      0.526903     1.613450    -1.884600
C      1.215100     1.630930    -1.038180

methane125
C      0.977714     1.118600    -0.264095
H     -0.091029     1.000000     0.000028
H      1.242040     0.375991    -1.032250
H      1.584730     0.960757     0.638015
H      1.164080     2.132720    -0.641080

water166
O     10.000000    10.000000    10.000000
H      9.519900    10.680200     9.523960
H      9.382580     9.416830    10.444300

water89
O     10.000000    10.000000    10.000000
H     10.711900    10.155200    10.623100
H     10.358400     9.920500     9.114090

methanol20
H      1.409710     2.399070    -1.276340
O      0.918680     1.589960    -1.269070
H      0.651798     1.641390     0.784907
H      0.456471     0.052686     0.028718
H      2.075150     0.744842     0.225517
C      1.032440     0.980422     0.004686

methanol14
H     -0.558663     0.654280    -1.072300
O      0.040334     0.133668    -0.555574
H      1.586030     1.510730    -0.773833
H      0.587710     1.718850     0.672737
H      1.741270     0.381208     0.583768
C      1.032440     0.980422     0.004686

water38
O     10.000000    10.000000    10.000000
H     10.711900    10.155200    10.623100
H     10.358400     9.920500     9.114090

formaldehyde67
O      1.013160     0.999902    -0.000000
H      0.907445    -0.266769     1.596210
H      1.792820     1.321710     1.854150
C      1.240530     0.679578     1.167220

formaldehyde168
O      1.013160     0.999902    -0.000000
H      2.381280    -0.070574     1.070410
H      2.085180     1.663620     1.600810
C      1.839050     0.861341     0.903690

methane68
H      0.315909     1.148670    -0.839374
H      1.865090     1.656890    -0.112718
H      1.332750    -0.037827     0.025121
H      0.474670     1.226870     0.928032
C      0.999002     0.999952     0.000058

\end{verbatim}

\pagebreak
\section{Coordinates for test-set molecules}\label{section:cords4}
\begin{verbatim}
ethane140
C      0.704789     0.169311    -0.199182
C     -0.336198     0.999753     0.000194
H      1.585560     0.505960    -0.730982
H      0.703859    -0.844512     0.178461
H     -0.311037     2.015010    -0.375195
H     -1.212730     0.678073     0.541440

ethylene53
C     -1.447600     0.497825    -0.554077
C     -0.336198     0.999753     0.000194
H     -2.403930     0.596751    -0.057231
H     -1.424300    -0.012789    -1.513650
H     -0.360687     1.494180     0.962482
H      0.616265     0.915950    -0.509468

methane173
C      0.987299     1.003440     0.190548
H     -0.091029     1.000000     0.000028
H      1.517630     1.316460    -0.710630
H      1.347350     0.008327     0.438848
H      1.222720     1.689410     1.016310

formaldehyde18
O      1.013160     0.999902    -0.000000
H      2.574610     1.374200     1.255930
H      1.373560     0.092417     1.790070
C      1.663860     0.818947     1.029940

methane104
C      0.926069     1.092350    -0.392636
H     -0.091029     1.000000     0.000028
H      1.015970     0.547372    -1.328530
H      1.639710     0.720018     0.347314
H      1.148540     2.135610    -0.584572

methane17
H      1.010120    -0.093152    -0.037683
H      1.825290     1.409520    -0.579551
H      0.052868     1.365970    -0.408566
H      1.107470     1.313660     1.039140
C      0.999002     0.999952     0.000058

ammonia144
H      1.125740     0.495018    -0.859590
H      1.563270     0.412701     0.692552
H      0.031839     0.772016     0.305215
N      0.995331     0.939970     0.039952

water120
O     10.000000    10.000000    10.000000
H     10.359000     9.276610     9.483330
H     10.645000    10.282100    10.650700

water35
O     10.000000    10.000000    10.000000
H      9.613060    10.052900    10.874500
H      9.515290    10.569000     9.401170

methanol30
H      0.709476     1.842350    -0.582889
H      1.946420     0.582476    -0.440952
O      0.032467    -0.027329     0.019979
H     -0.756401     0.317865     0.412981
H      1.263700     1.316950     1.017510
C      1.032440     0.980422     0.004686

methanol45
H      1.264390     1.237090    -1.030340
H     -0.049813     0.869274     0.095120
O      1.500940     1.984470     0.889560
H      2.437900     2.076840     0.796041
H      1.492060     0.018135     0.235031
C      1.032440     0.980422     0.004686

ethylene100
C      0.607156     1.266710     0.916643
C     -0.336198     0.999753     0.000194
H      1.173750     0.439645     1.392550
H      0.878961     2.304580     1.207250
H     -0.577840    -0.013253    -0.265105
H     -0.885468     1.797150    -0.475380

ammonia19
H      0.445872     0.830841    -0.824314
H      1.666940     0.224161     0.048049
H      0.378501     0.859092     0.792353
N      1.002530     0.984772     0.006126

ammonia123
H      1.056040     0.577922    -0.891603
H      0.639458     1.908700    -0.068977
H      1.937280     1.045650     0.376881
N      0.987155     0.962389     0.039316

methane60
H      1.023880     0.789414    -1.070270
H      0.923417     0.064942     0.560522
H      1.916110     1.523540     0.281343
H      0.132935     1.628990     0.216577
C      0.999002     0.999952     0.000058

ethane108
C     -1.230010     0.349915     0.750149
C     -0.336198     0.999753     0.000194
H     -1.477180     0.679722     1.760490
H     -1.740060    -0.541145     0.375924
H      0.151763     1.889430     0.385279
H     -0.083601     0.669757    -1.000270

ammonia120
H      0.753137     0.633410    -0.885941
H      1.994170     0.973531     0.090420
H      0.678411     0.251272     0.690159
N      0.984233     0.965407     0.040724

formaldehyde120
O      1.013160     0.999902    -0.000000
H      2.221860     2.179880    -1.140050
H      0.673273     1.580450    -1.924270
C      1.307230     1.595420    -1.036530

methane134
C     -0.537195     1.121540     0.991142
H     -0.091029     1.000000     0.000028
H     -0.788475     0.142749     1.402700
H     -1.444350     1.726020     0.925262
H      0.169570     1.615630     1.659380

ammonia166
H      0.936428     0.671831    -0.940282
H      1.955800     0.694686     0.318631
H      0.402570     0.296354     0.535961
N      1.013450     0.927154     0.034964

ethylene98
C      0.386274     1.830350    -0.771786
C     -0.336198     0.999753     0.000194
H      1.356160     2.141930    -0.450913
H      0.030189     2.204110    -1.708010
H      0.037415     0.644848     0.934084
H     -1.292640     0.702852    -0.323258

methane94
H      1.156080     0.033239    -0.469733
H      0.912391     1.774780    -0.767629
H      1.850290     1.219720     0.648722
H      0.080261     0.968424     0.590696
C      0.999002     0.999952     0.000058

ammonia133
H      1.443040     0.706455    -0.836028
H      1.585220     0.591062     0.767803
H      0.136661     0.416840     0.077077
N      0.991859     0.956923     0.033511

ethane55
H     -0.431944    -0.262536    -1.080550
H      0.963953    -1.141430    -0.466616
C      0.261669    -0.337825    -0.241785
C      0.998505     0.999338    -0.001117
H      1.570210     1.294960    -0.881848
H     -0.312791    -0.637648     0.636342
H      1.691860     0.922415     0.837393
H      0.293276     1.801140     0.221223

ammonia192
H      0.173770     1.485980    -0.004216
H      1.190880     0.648161    -0.938802
H      1.769950     1.555960     0.267097
N      1.017450     0.929087     0.016282

methane145
C     -1.148140     0.860485    -0.265919
H     -0.091029     1.000000     0.000028
H     -1.584230     0.063148     0.330727
H     -1.231350     0.595481    -1.317490
H     -1.704700     1.772610    -0.071931

formaldehyde91
O      1.013160     0.999902    -0.000000
H      1.841200     0.145586     1.655490
C      0.997428     0.647278     1.179630
H      0.138511     0.812795     1.831230

ammonia141
H      1.533780     1.006480    -0.813591
H      0.366616     0.156889    -0.076389
H      1.646540     0.675554     0.763094
N      0.993050     0.942894     0.038566

ammonia104
H      1.316400     0.984268    -0.917901
H      1.811110     0.855243     0.616299
H      0.438070     0.128887     0.126480
N      0.990208     0.971536     0.035719

formaldehyde118
O      1.013160     0.999902    -0.000000
H      1.470060    -0.168380     1.606280
H      0.232042     1.161020     1.872560
C      0.904109     0.658661     1.176560

ethane163
C     -0.334738     1.968900    -0.958671
C     -0.336198     0.999753     0.000194
H     -1.251700     2.410210    -1.344360
H      0.595301     2.327400    -1.371290
H     -1.283870     0.658816     0.393895
H      0.570992     0.552515     0.395057

ammonia147
H      0.591802     0.879984    -0.883401
H      2.008470     0.968414    -0.100740
H      0.738938     1.840060     0.408068
N      1.005800     0.935673     0.038246

formaldehyde66
O      1.013160     0.999902    -0.000000
H      2.711380     1.293290     1.087550
H      1.413510     0.238573     1.848340
C      1.723240     0.841301     0.993394

formaldehyde163
O      1.013160     0.999902    -0.000000
H      2.983080     0.454338     0.069570
H      2.612150     2.252760     0.139499
C      2.221600     1.236660     0.070465

formaldehyde180
O      1.013160     0.999902    -0.000000
H      1.815850     1.444510     1.821200
C      1.857780     0.880859     0.888742
H      2.706500     0.200836     0.805972

ethane57
H     -0.242613    -0.117461    -1.416540
H      1.293320    -0.897207    -1.051590
C      0.478361    -0.313524    -0.621577
H      1.492200     1.619260    -0.750776
C      0.998505     0.999338    -0.001117
H     -0.014894    -0.935663     0.126625
H      1.719370     0.799660     0.792688
H      0.182538     1.581800     0.428393

methane109
C      0.670457     1.189050     0.780541
H     -0.091029     1.000000     0.000028
H      1.586580     1.559160     0.317138
H      0.891375     0.275554     1.328530
H      0.287312     1.927440     1.485460

water71
O     10.000000    10.000000    10.000000
H      9.647260    10.511600    10.728300
H      9.525660    10.215500     9.194270

water170
O     10.000000    10.000000    10.000000
H     10.626200    10.407100    10.599600
H      9.663180    10.663700     9.395720

ethylene55
C     -0.015305     1.391250     1.246620
C     -0.336198     0.999753     0.000194
H     -0.453294     0.926060     2.116740
H      0.691975     2.195680     1.408370
H     -1.043320     0.196253    -0.163210
H      0.101050     1.472860    -0.869200

methane181
C      0.957657     0.663038    -0.027311
H     -0.091029     1.000000     0.000028
H      1.475850     1.111070     0.824132
H      1.451220     0.981967    -0.949642
H      1.034690    -0.423166     0.048236

formaldehyde173
O      1.013160     0.999902    -0.000000
H     -0.352013    -0.108431     1.030990
H     -0.878172     1.567770     0.497718
C     -0.088155     0.816136     0.516420

formaldehyde154
O      1.013160     0.999902    -0.000000
H      2.199770     1.095490     1.653200
H      0.676030     0.078717     1.790870
C      1.299400     0.719898     1.165590

methanol5
H      0.737850     0.345300    -0.828430
O      1.540850     2.221060    -0.477540
H      0.148290     1.155780     0.629938
H      1.774440     0.458020     0.607998
H      1.744250     2.781090     0.249556
C      1.032440     0.980422     0.004686

formaldehyde55
O      1.013160     0.999902    -0.000000
C      2.116030     1.096750     0.537128
H      2.399830     0.509014     1.411010
H      2.886920     1.778450     0.176285

ammonia98
N      1.008520     0.975287    -0.008309
H      0.393360     0.575845    -0.706959
H      1.562290     1.681420    -0.480198
H      0.409364     1.474610     0.638881

methane85
H      1.601680     1.363670    -0.836743
H      0.249946     0.290590    -0.356877
H      1.651120     0.505930     0.722997
H      0.497031     1.842730     0.473047
C      0.999002     0.999952     0.000058

methane26
H      0.829864     0.972392    -1.079270
H      2.000300     0.621809     0.225583
H      0.249574     0.371072     0.490306
H      0.907267     2.028330     0.357128
C      0.999002     0.999952     0.000058

methane92
H      1.463970     0.122345    -0.448969
H      0.747869     1.723300    -0.778175
H      0.089472     0.693849     0.522536
H      1.695860     1.451800     0.705586
C      0.999002     0.999952     0.000058

methanol72
H      0.675046     0.655293    -0.974740
H     -0.554545     0.445962     1.005210
H      0.858186     2.054780     0.097300
H      2.108690     0.807818     0.057322
O      0.377041     0.278863     1.050490
C      1.032440     0.980422     0.004686

\end{verbatim}